\documentclass{aa}
\usepackage{txfonts}
\usepackage{graphicx}
\usepackage{natbib}
\usepackage{supertabular}
\usepackage{longtable}
\bibpunct{(}{)}{;}{a}{}{,} 
\begin{document}
\title{Chemistry of heavy elements in the Dark Ages}
\titlerunning{Chemistry of heavy elements in the Dark Ages}

\author{P. Vonlanthen \inst{1} \and T. Rauscher \inst{2} \and C. Winteler \inst{2} \and D. Puy \inst{1} \and M. Signore \inst{3} \and V. Dubrovich \inst{4}}

\institute{Universit\'e Montpellier II - GRAAL, CNRS - UMR 5024, place Eug\`ene Bataillon, 34095 Montpellier, France \and 
Department of Physics, University of Basel, 4056 Basel, Switzerland \and LERMA, Observatoire de Paris, 75014 Paris, France \and SPb Branch of Special Astrophysical Observatory, RAS, St Petersburg, Russia\\
\email{Patrick.Vonlanthen@graal.univ-montp2.fr}
}

\date{Received / Accepted}

\abstract
{Primordial molecules were formed during the Dark Ages, i.e. the time between recombination and reionization in the early Universe. They were the constitutents of the first proto-stellar clouds. Standard Big Bang nucleosynthesis predicts the abundances of hydrogen, helium, lithium, beryllium, and their isotopes in the early Universe. Heavier nuclei such as carbon, nitrogen, or oxygen are formed in trace amounts only. In non-standard Big Bang nucleosynthesis models, it is possible to synthesize larger quantities of these heavier elements. The latter are interesting because they can form molecules with a high electric dipole moment which can increase the cooling in collapsing protostellar structures.
}
{The purpose of this article is to analyze the formation of primordial molecules based on heavy elements during the Dark Ages, with elemental abundances taken from different nucleosynthesis models.
}
{We present calculations of the full non-linear equation set governing the primordial chemistry. We consider the evolution of 45 chemical species and use an implicit multistep method of variable order of precision with an adaptive stepsize control.
}
{For the first time the cosmological recombination of heavy elements is presented. We find that the most abundant Dark Ages molecules based on heavy elements are CH and OH. When considering initial conditions given by the standard Big Bang nucleosynthesis model, we obtain relative abundances $\mathrm{[CH]} = n_{\mathrm{CH}}/n_{\mathrm{b}} = 6.2 \times 10^{-21}$ and $\mathrm{[OH]} = n_{\mathrm{OH}}/n_{\mathrm{b}} = 1.2 \times 10^{-23}$ at $z = 10$, where $n_{\mathrm{b}}$ is the total number density. But non-standard nucleosynthesis can lead to higher heavy element abundances while still satisfying the observed primordial light abundances. In that case, we show that the abundances of molecular species based on C, N, O and F can be enhanced by two orders of magnitude, leading to a CH relative abundance higher than that of HD$^+$ or H$_2$D$^+$.
}
{}


\maketitle

\section{Introduction}
The formation of light elements is a central problem of modern cosmology. Big Bang nucleosynthesis (BBN) provides an important testing ground for the physics of the primordial Universe. The proceedings of nucleosynthesis are relatively well understood. Nucleosynthesis is connected with two important cosmological events. The first one is the freeze-out, at T $\approx$ 0.8 MeV, of the weak interactions which interconvert neutrons and protons and thus setting the initial proton-to-neutron ratio for nucleosynthesis. The second event is the annihilation of thermal e$^-$-e$^+$ pairs in the temperature range 1 MeV - 20 keV, delaying the onset of nucleosynthesis by releasing additional heat but eliminating the possibility of positron captures.

Thereafter, thermal fusion reactions in the plasma first produce deuterium nuclei:
\begin{equation}
\mathrm{p} + \mathrm{n} \rightarrow \mathrm{D} + \gamma,
\end{equation}
and then helium (\element[][3]{He}, \element[][4]{He}), lithium (\element[][7]{Li}) and beryllium (\element[][7]{Be}). Cosmological expansion plays a crucial role during the whole process and determines the timescale for the nucleosynthesis during the cooling of the Universe.
Observations agree well with the predicted abundances of the standard BBN model. See for example \citet{bur01}, but see also \citet{ste07} for the latest observations (and in particular his Fig. 13).

Due to the proton-rich freeze-out from high temperature, standard BBN predicts the highest abundance for the nucleus with the largest binding energy, which is $^4$He. Beyond $^4$He, charged particle reactions are inefficient due to the rising Coulomb barrier and the small density. Therefore, only tiny amounts of heavier nuclei are produced in standard BBN, if any. However, a large number of extensions to the standard BBN have been suggested, among which we find Non-Standard Big Bang Nucleosynthesis (hereafter NSBBN) scenarios considering density fluctuations in the early Universe. Some of these models show increased abundances of heavy elements (see e.g.\ \citealt{dol93,jed94,rau94,khl07} and references therein). The original motivation for such studies was to obtain an average baryon density equal to the critical density for a flat universe and to simultaneously satisfy the constraints given by primordial abundance observations. While this was shown to be unfeasible (and unnecessary), density fluctuations are present in the cosmic microwave background. The density fluctuations affecting BBN are of different magnitude, though, but many different mechanisms have been suggested, acting during the inflationary phase, during baryogenesis or later. Thus, NSBBN with density fluctuations remains an interesting model for obtaining primordial heavy elements.

During the recombination period the nuclei became progressively neutralized, which led to molecular formation. However, at early epochs, where a total absence of dust grains appears justified, the chemistry is different from the typical interstellar medium astrochemistry.

Several groups proposed an assembled, comprehensive set of reactions for the early Universe \citep{lep84,puy93,GP98,sta98,sig99,lep02,pfe03,puysig07,gloabel08,glosav08}. The chemical network is coupled with the matter and radiation temperatures and with the matter density. Primordial chemistry of heavy nuclei has been poorly studied, although \citet{lip07} analyzed the possibility to detect rotational lines of primordial CH, and \citet{puy07} analyzed the possibility to form molecular fluorine HF in the early Universe.

In this paper we will discuss the chemistry of eight elements (hydrogen, deuterium, helium, lithium, carbon, nitrogen, oxygen and fluorine) in the early Universe, for two different contexts: the standard Big Bang nucleosynthesis model and non-standard nucleosynthesis based on baryon density fluctuations. Heavier elements are particularly interesting because they can form molecules with a high electric dipole moment. Even if their abundances are proven to be quite low, they could still furnish an interesting cooling agent in the gravitational collapse of protostructures, as molecular abundances are prone to drastic variations in such environments. In Sect. \ref{sect:bbn}, we summarize the standard and non-standard BBN models we used and the predicted abundances. We describe and calculate, in Sect. \ref{sect:chemistry}, the mechanisms of recombination and primordial chemistry which act for the eight elements H, D, He, Li, C, N, O and F. In Sect. \ref{sect:concl} we discuss our results.

\section{Big Bang nucleosynthesis}
\label{sect:bbn}
\subsection{Standard model of primordial nucleosynthesis}
Standard BBN assumes a homogeneous baryon density throughout the expanding and cooling Universe. The abundance ratio of protons and neutrons is set by the freeze-out of the weak interaction and subsequent decay of the neutrons until the onset of nucleosynthesis. At initially high temperature a nuclear statistical equilibrium is established, favoring the
formation of the strongly bound $^4$He as soon as the photodisintegrations cease. All neutrons are consumed but a large fraction of protons remains because of the higher initial proton abundance. Due to the high entropy, i.e.\ low density, the triple-$\alpha$ reaction is inefficient and the formation of elements beyond He is suppressed. For this reason, there are few accounts of standard BBN heavy element abundances in literature.

Here, we will consider BBN results by \citet{winteler07}, obtained with a code based on a modified version of the Basel network code. Modifications include a consistent evolution of temperature and density in the early universe, including weak freeze-out, based on methods by \citet{kawano92}, and an improved reaction network with updated reaction rates, for details see \citet{winteler07}. The abundances were calculated for $\eta = 6.22 \times 10^{-10}$, in accordance with the latest result of the Wilkinson Microwave Anisotropy Probe (WMAP) experiment ($\Omega_{\mathrm{b}} h^2 = 0.02273 \pm 0.00062$, giving $\eta_{10} = 6.225 \pm 0.170$, see \citealt{dun09}). The weak freeze-out process was carefully followed until the nucleosynthesis phase, thus setting the initial neutron and proton abundances consistently. The first column of Table \ref{mfrac} gives the relative abundances for this SBBN case.
Here and throughout this paper, we consider relative abundances for species $\xi$ such as:
\begin{equation}
[\xi] = n_{\xi}/n_{\mathrm{b}},
\end{equation}
where $n_{\xi}$ is the number density of species $\xi$ and $n_{\mathrm{b}}$ the total number density. These values are the initial conditions for our calculations of the Standard Big Bang Chemistry (hereafter SBBC). The relative abundances for the light nuclei H, D and He are in accordance with the observational constraints \citep[see][]{bur01,cyb03,ste07}, while lithium seems to be more abundant than deduced from observations. Indeed, \citet{ste07} adopts the following primordial abundances\footnote{Alternative abundances for \element[][4]{He} and \element[][7]{Li} are written in parentheses.}: (D/H)$_{\mathrm{P}} = 2.68^{+0.27}_{-0.25} \times 10^{-5}$, Y$_{\mathrm{P}} = 0.240 \pm 0.006 \; (< 0.251 \pm 0.002)$ and 12 + $\log$(Li/H)$_{\mathrm{P}} = 2.1 \pm 0.1 \; (2.5 \pm 0.1)$, while the values of first column of Table \ref{mfrac} give, after a straightforward calculation: (D/H)$_{\mathrm{P}}$ $= 2.35 \times 10^{-5}$, Y$_{\mathrm{P}} = 0.246$ and 12 + $\log$(Li/H)$_{\mathrm{P}}$ = 3.3. Let us note that lithium has been observed only in the absorption spectra of very old, very metal-poor stars (Population II stars); they are considered as ideal for probing the primordial abundance of lithium. But lithium is a fragile nucleus which is easily destroyed in the interiors of stars. Therefore, if one admits a large primordial value for lithium, one must also assume a large and uniforme depletion in stars, over a range of stellar masses.

In Table \ref{mfrac} we also quote the amounts of the elements C to F produced in this standard BBN, as they are included in the SBBC calculations.
\begin{table}[!h]
\caption{Relative abundances $\left[\xi\right]=n_{\xi}/n_{b}$ of the elements
at the end of Big Bang nucleosynthesis for the SBBN and for two non-standard BBN scenarios.}
\label{mfrac}
\centering
\begin{tabular}{cccc}
\hline
\hline 
 & SBBN & \multicolumn{2}{c}{NSBBN}\tabularnewline
 &  & $\left(f_\mathrm{v}=0.8;\, R=10\right)$ & $\left(f_\mathrm{v}=10^{-5};\, R=1000\right)$\tabularnewline
\hline
$\left[\mathrm{H}\right]$ & $0.889$ & $0.889$ & $0.888$\tabularnewline
$\left[\mathrm{D}\right]$ & $2.092\times10^{-5}$ & $2.45\times10^{-5}$ & $2.13\times10^{-5}$\tabularnewline
$\left[\mathrm{He}\right]$ & $0.111$ & $0.111$ & $0.112$\tabularnewline
$\left[\mathrm{Li}\right]$ & $1.77\times10^{-9}$ & $1.97\times10^{-9}$ & $2.30\times10^{-9}$\tabularnewline
$\left[\mathrm{C}\right]$ & $2.51\times10^{-15}$ & $4.00\times10^{-15}$ & $8.45\times10^{-14}$\tabularnewline
$\left[\mathrm{N}\right]$ & $2.32\times10^{-16}$ & $2.46\times10^{-16}$ & $3.44\times10^{-14}$\tabularnewline
$\left[\mathrm{O}\right]$ & $3.22\times10^{-19}$ & $3.37\times10^{-19}$ & $8.20\times10^{-17}$\tabularnewline
$\left[\mathrm{F}\right]$ & $3.28\times10^{-27}$ & $3.61\times10^{-27}$ & $1.63\times10^{-24}$\tabularnewline
\hline
\end{tabular}
\end{table}

\subsection{Non-Standard Big Bang Nucleosynthesis}
\label{nsbbn}
While heavy elements are not produced at high levels in SBBN, BBN models assuming density fluctuations allow for different baryon densities $\rho_{1,2,...}$ in different zones, leading to altered nucleosynthesis. A large number of
possibilities for creating small scale density perturbations in the very early Universe have been suggested
in literature \citep{aff85,applegate85,malaney88,dol93,mat04,khl07}. Here, we are not focussing on a specific origin but just assume the occurrence of such fluctuations
and use the geometry as open parameter. As customary, we apply a two-zone model where the densities $\rho_1$ and $\rho_2$ of the zones are given by the density ratio $R=\rho_1/\rho_2=\eta^{(1)}/\eta^{(2)}$, the volume fraction $0\leq f_\mathrm{v}\leq 1$ of zone 1, and the additional constraint that the averaged density has to reproduce the WMAP value (as in the SBBN): $\eta_{10}^\mathrm{WMAP}=6.22=\eta^{(1)}f_\mathrm{v}+\eta^{(2)}(1-f_\mathrm{v})$. This leaves two open parameters, $R$ and $f_\mathrm{v}$, but we are further limited by the observed primordial light abundances. Assuming that the Li abundance is only a weak constraint because of the complicated stellar
depletion of Li, several regions in the $R$,$f_\mathrm{v}$-space remain allowed.

To study the impact of elevated levels of heavy nuclei, we used the same code by \citet{winteler07} as for the SBBN but followed nucleosynthesis in zones promising for synthesizing heavy elements. Table \ref{mfrac} shows the results for two models, representative for two extreme cases. The abundances shown in the table are already the final abundances, mixed from the two zones in each model:
\begin{equation}
\left[\xi\right]=\frac{f_\mathrm{v}\eta^{(1)}\left[\xi\right]^{(1)}+(1-f_\mathrm{v})\eta^{(2)}\left[\xi\right]^{(2)}}{\eta_{10}^\mathrm{WMAP}} \quad.
\end{equation}

The first case ($f_\mathrm{v}=0.8$, $R=10$) is similar to the scenario
studied in \citet{rau94}. The density of $\rho_1$ has to stay close to the global $\eta_{10}^\mathrm{WMAP}$, while $\rho_2$ is ten times lower. Following \citet{rau94}, to maximize the production of heavy nuclei we assume complete diffusion of the uncharged neutrons out of the high-density region. The electrically charged protons remain trapped in that region because of their much shorter mean free path. Therefore, baryon density inhomogeneities become local variations of the neutron-to-proton ratio, with free neutrons left over in the low-density zone after the initial formation of $^4$He. Neutron captures can then produce neutron-rich isotopes, allowing to bypass the slow triple-$\alpha$ reaction and leading to heavy elements. Our results are consistent with previous results from literature. When obtaining H, D, $^4$He abundances close to the observed ones, the $^7$Li abundance becomes even higher than in the standard model. At the same time, the heavy element abundances remain at low levels, only very slightly higher than for the SBBN. This was already pointed out in \citet{rau94}, where it was also found that the light element constraints prevent considerable formation of heavy nuclei.

The second case investigated here ($f_\mathrm{v}=10^{-5}$, $R=1000$) comprises the other end of the allowed spectrum, tiny pockets of extremely high density embedded in a background with almost standard density. This is similar to the scenario originally introduced by \citet{jed94} and later also discussed by \citet{mat04,mat05,mat07}. In the high density pocket the path to heavy elements is opened by an efficient triple-$\alpha$ reaction. At the same time, increased destruction of $^7$Li keeps its abundance low. This simultaneously allows to achieve light element abundances compatible with those of the SBBN and to increase the heavy element production considerably. However, due to the small volume fraction of the high density zone (which is necessary to obtain the proper global baryon density) the final heavy element abundances after complete mixing remain small, although two orders of magnitude higher than for the standard case.

In the following, we will use the abundance values from both NSBBN scenarios as initial conditions for our calculations of the Non-Standard Big Bang Chemistry (hereafter NSBBC1 and NSBBC2 respectively).

\section{Dark Ages chemistry}
\label{sect:chemistry}
\subsection{Equations of evolution}
The primordial gas is a mixture of hydrogen, deuterium, helium, lithium, \ldots $ \ $ so there are many possibilities of reactions. Change in the number density of a given chemical species due to chemical reactions depends on the densities of the species involved in these reactions and on the reaction rates, which themselves depend on the matter and radiation temperatures. But of course cosmic expansion also plays a crucial role. Thus, it is necessary to take into account the following set of differential equations (see for example \citealt{puy93}) in the context of the expanding Universe, characterized by the scale factor $a$:
\begin{eqnarray}
\frac{\mathrm{d}T_{\mathrm{r}}}{\mathrm{d}t} & = & - \frac{1}{a} \frac{\mathrm{d}a}{\mathrm{d}t} T_{\mathrm{r}} \label{Tr}\\
\frac{\mathrm{d}T_{\mathrm{m}}}{\mathrm{d}t} & = & - 2 \frac{1}{a} \frac{\mathrm{d}a}{\mathrm{d}t} T_{\mathrm{m}} + \frac{8}{3} \frac{\sigma_{\mathrm{T}} a_{\mathrm{r}}}{m_{\mathrm{e}} c} T_{\mathrm{r}}^{4} \left( T_{\mathrm{r}} - T_{\mathrm{m}} \right) \frac{n_{\mathrm{e}}}{n_{\rm{b}}} - T_{\mathrm{m}} \frac{1}{n_{\rm{b}}} \frac{\mathrm{d}n_{\rm{b}}}{\mathrm{d}t}
\label{Tm}\\
\frac{\mathrm{d}n_{\rm{b}}}{\mathrm{d}t} & = & - 3 \frac{1}{a} \frac{\mathrm{d}a}{\mathrm{d}t} n_{\rm{b}} - \sum_{\xi} \left( \frac{\mathrm{d}n_{\xi}}{\mathrm{d}t} \right)_{\rm{chem}} \label{nb}\\
\frac{\mathrm{d}n_{\xi}}{\mathrm{d}t} & = & - 3 \frac{1}{a} \frac{\mathrm{d}a}{\mathrm{d}t} n_{\xi} + \left( \frac{\mathrm{d}n_{\xi}}{\mathrm{d}t} \right)_{\rm{chem}} \label{nxi}
\end{eqnarray}
In these equations, $T_{\mathrm{r}}$ is the radiation temperature, $T_{\mathrm{m}}$ the gas temperature, $n_{\rm{b}}$ the total number density and $n_{\xi}$ the number density of species $\xi$. The right side of equation (\ref{Tr}) and the first term of the right side of equation (\ref{Tm}) represent the decrease of $T_{\mathrm{r}}$ and $T_{\mathrm{m}}$ due to the expansion. The second term of the right hand side of (\ref{Tm}) is the energy transfer from radiation to the gas via Compton diffusion of the CMB photons on the electrons \citep{komp57,pee68}. In this term, $\sigma_{\mathrm{T}}$ is the Thomson cross section, $a_{\mathrm{r}}$ the radiation constant, $m_{\mathrm{e}}$ the electronic mass, $c$ the speed of light and $n_{\mathrm{e}}$ the electron number density. Note that we do not consider the energy transfer between gas and radiation via molecular heating and cooling functions, since it has been shown that this contribution to the evolution of the gas temperature is negligible \citep{puy96,puy97}. In equations (\ref{nb}) and (\ref{nxi}), the first term of the right hand side characterizes again the density decrease due to cosmic expansion, while the second term is the contribution of chemical reactions:
\begin{equation}
\left( \frac{\mathrm{d}n_{\xi}}{\mathrm{d}t} \right)_{\rm{chem}} = \sum_{\xi_1 \xi_2} k_{\xi_1 \xi_2} n_{\xi_1} n_{\xi_2} - \sum_{\xi'} k_{\xi \xi'} n_{\xi} n_{\xi'}.
\label{cin}
\end{equation}
$k_{\xi_1 \xi_2}$ is the rate of the reaction with reactants $\xi_1$ and $\xi_2$.

This system of ordinary differential equations governing the chemical abundances is stiff. We use an implicit multistep method of variable order of precision with an adaptive stepsize control. This method has excellent stability properties and is widely used for solving chemical kinetic problems \citep{hin95}. We solve the set of equation (\ref{Tr}) to (\ref{nxi}) from the initial redshift $z_{\rm init} = 10^4$, when the Universe was still totally ionized, to the final redshift $z_{\rm end} = 10$ (about the epoch of reionization).

Moreover, we consider the standard $\Lambda$CDM model, with Hubble parameter $H_0 = 71$ km s$^{-1}$ Mpc$^{-1}$, total matter density $\Omega_{\mathrm{m}} = 0.27$ (including dark matter density $\Omega_{\mathrm{DM}}$ = 0.226 and baryon density $\Omega_{\mathrm{b}}$ = 0.044) and dark energy density $\Omega_{\Lambda} = 0.73$ \citep{wmap5}.

\subsection{Cosmological recombination}
\label{recomb}
\begin{figure}
\vspace{0.5cm}
\begin{center}
\includegraphics[scale=0.30]{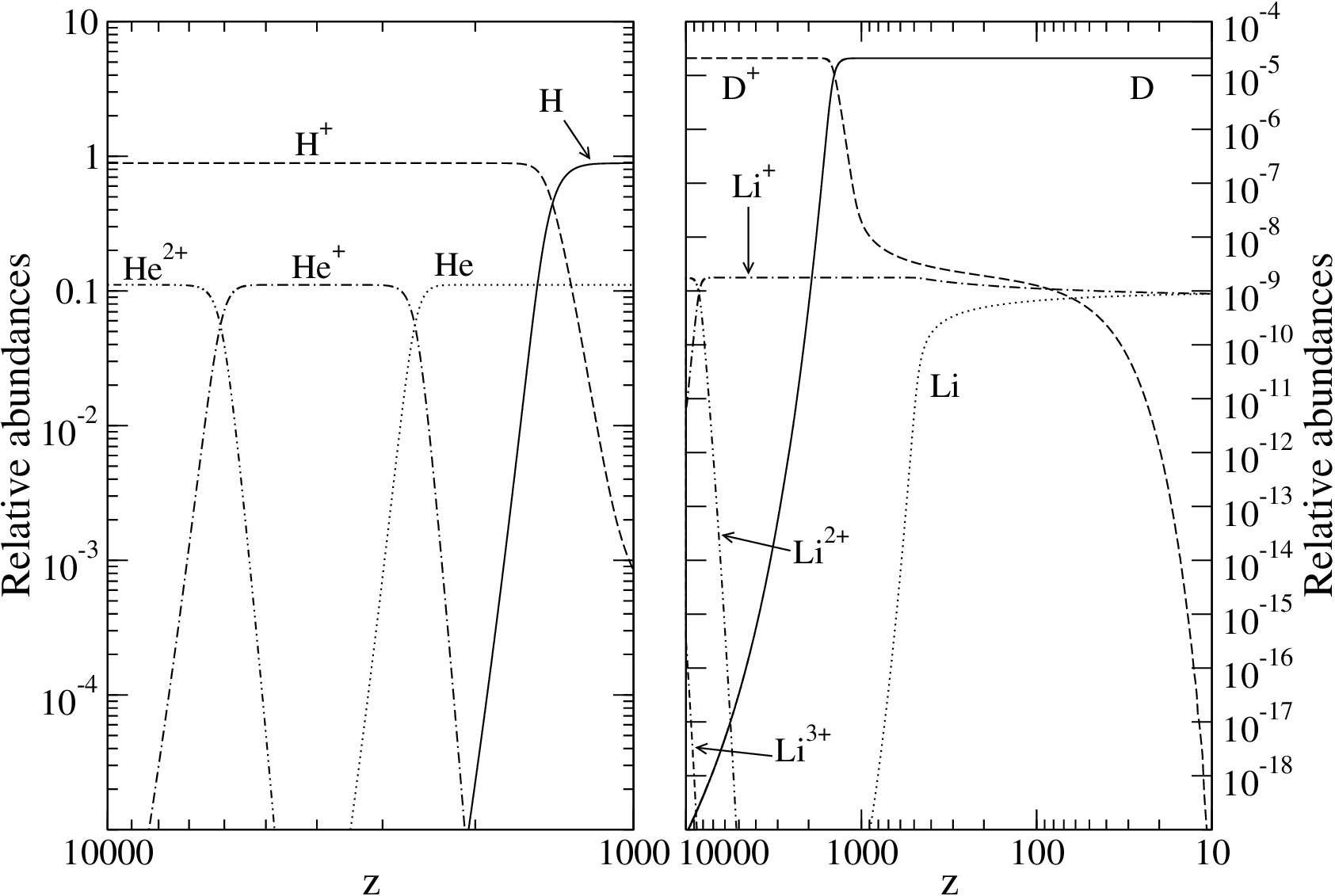}
\end{center}
\caption{Successive cosmological recombination for hydrogen, deuterium, helium and lithium in the standard model of primordial chemistry.}
\label{REC_HHeDLi}
\end{figure}

Primordial chemistry begins with the appearance of the first neutral species. Once a neutral species is formed, charge transfers with ions become efficient and lead to the formation of other neutral species and then molecular ions. Figure \ref{REC_HHeDLi} shows the evolution of the chemical abundances during the successive periods of recombination for the light elements H, D, He and Li in the standard model. Helium nuclei are the first to recombine because of their high electronic binding energy\footnote{New results on the complex helium and hydrogen recombinations are given in \citet{fendt08} (see also references therein).}. The recombination and photoionization rates for H and D are taken from \citet{abe97}. We use the rates of \citet{GP98} for He and for the reactions Li$^+$ + e$^-$ $\leftrightarrow$ Li + $\gamma$. Higher lithium recombinations and photoionisations are treated with the recombination rates of \citet{ver96} and the photoionisation cross sections of \citet{ver96a}. The recombination pattern for the two non-standard cases NSBBC1 and NSBBC2 are very similar, since the initial abundances of H, D, He and Li are almost identical to the SBBC model.
We define the recombination redshift $z_{\mathrm{rec}}$ as the redshift when the abundance of a given species equals the one of its corresponding ion. Table \ref{zrec} summarizes the different recombination redshifts for the standard calculation. Due to its low ionization energy and to the fact that charge transfer reactions between lithium and hydrogen remain active, lithium does not recombine totally, and both neutral and ionized Li tend to match their abundances (see Fig. \ref{REC_HHeDLi}).
\begin{table}[!h]
\caption{Recombination redshifts for hydrogen, deuterium and helium.}
\label{zrec}
\centering
\begin{tabular}{l l l l}
\hline
\hline
 & SBBC\\
\hline
$z_{\mathrm{rec}} \ (\mathrm{He^+})$ & 6102\\
$z_{\mathrm{rec}} \ (\mathrm{He})$ & 2604\\
$z_{\mathrm{rec}} \ (\mathrm{H}) = z_{\mathrm{rec}} \ (\mathrm{D})$ & 1425\\
\hline
\end{tabular}
\end{table}

\begin{figure}[!t]
\vspace{0.3cm}
\begin{center}
\includegraphics[scale=0.3]{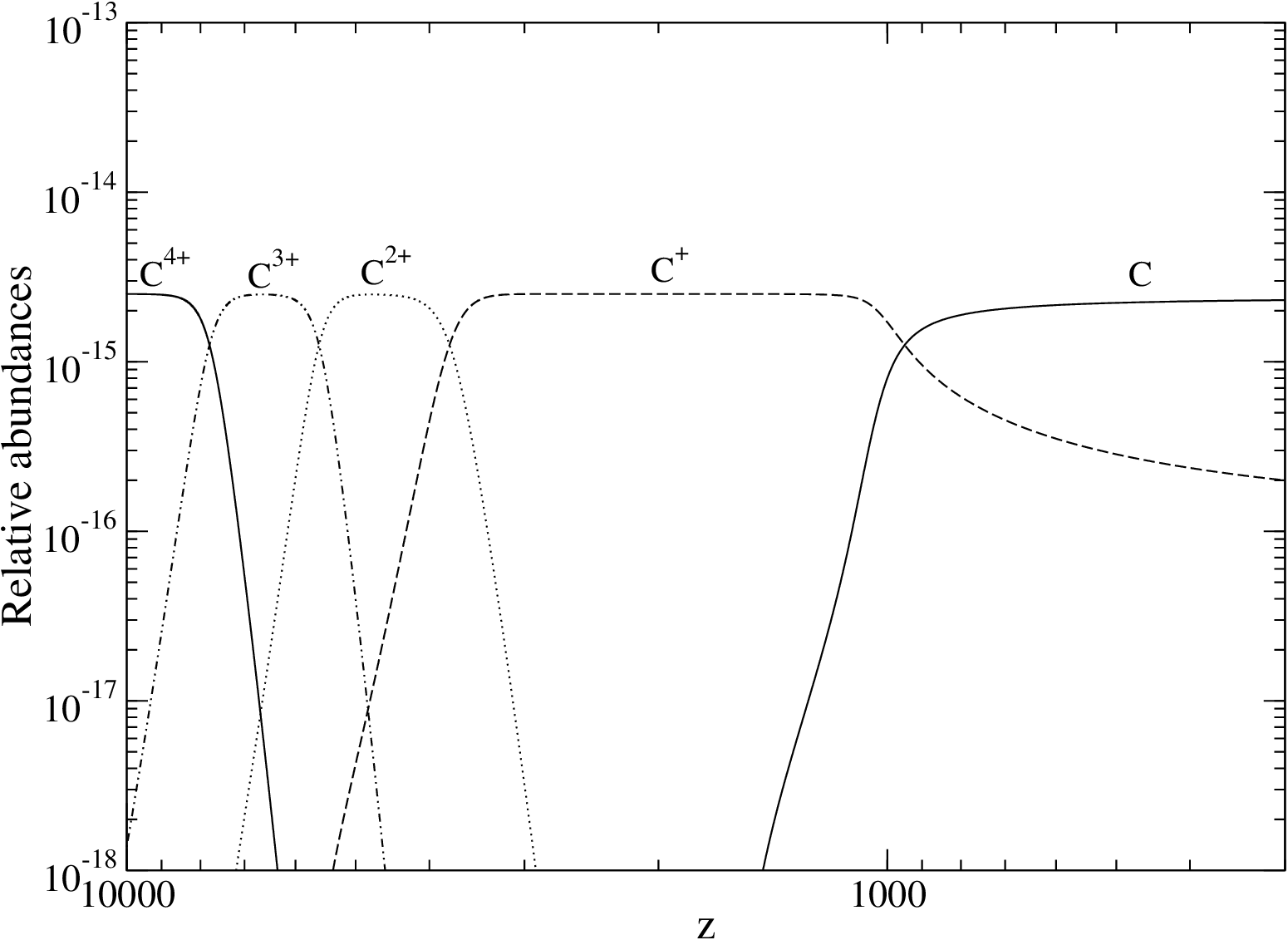}
\end{center}
\caption{Successive recombinations of the carbon ions in the standard model.}
\label{fig:recombic}
\end{figure}

\begin{figure}[!t]
\vspace{0.3cm}
\begin{center}
\includegraphics[scale=0.3]{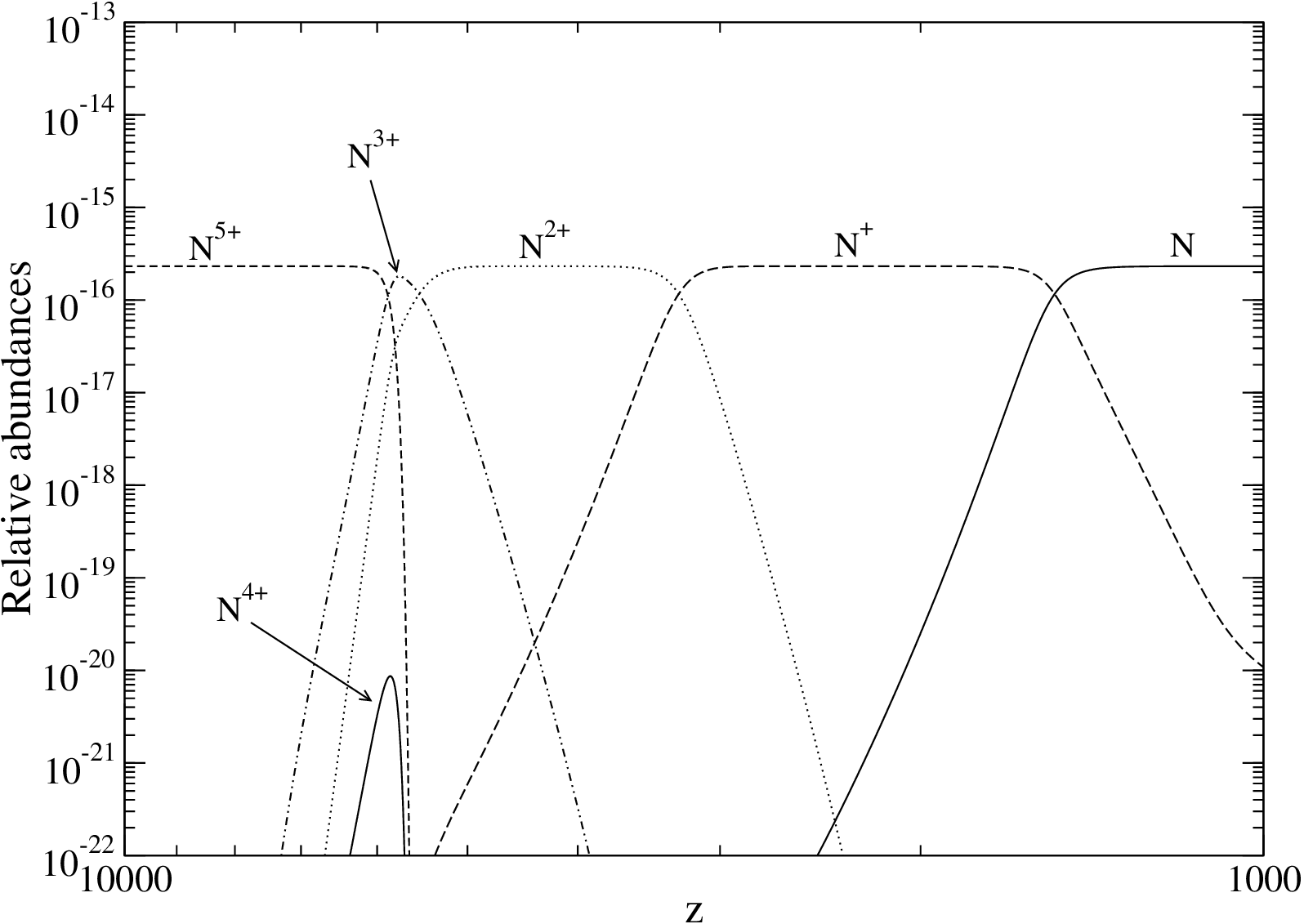}
\end{center}
\caption{Successive recombinations of the nitrogen ions in the standard model.}
\label{fig:recombin}
\end{figure}

\begin{figure}[!t]
\vspace{0.3cm}
\begin{center}
\includegraphics[scale=0.3]{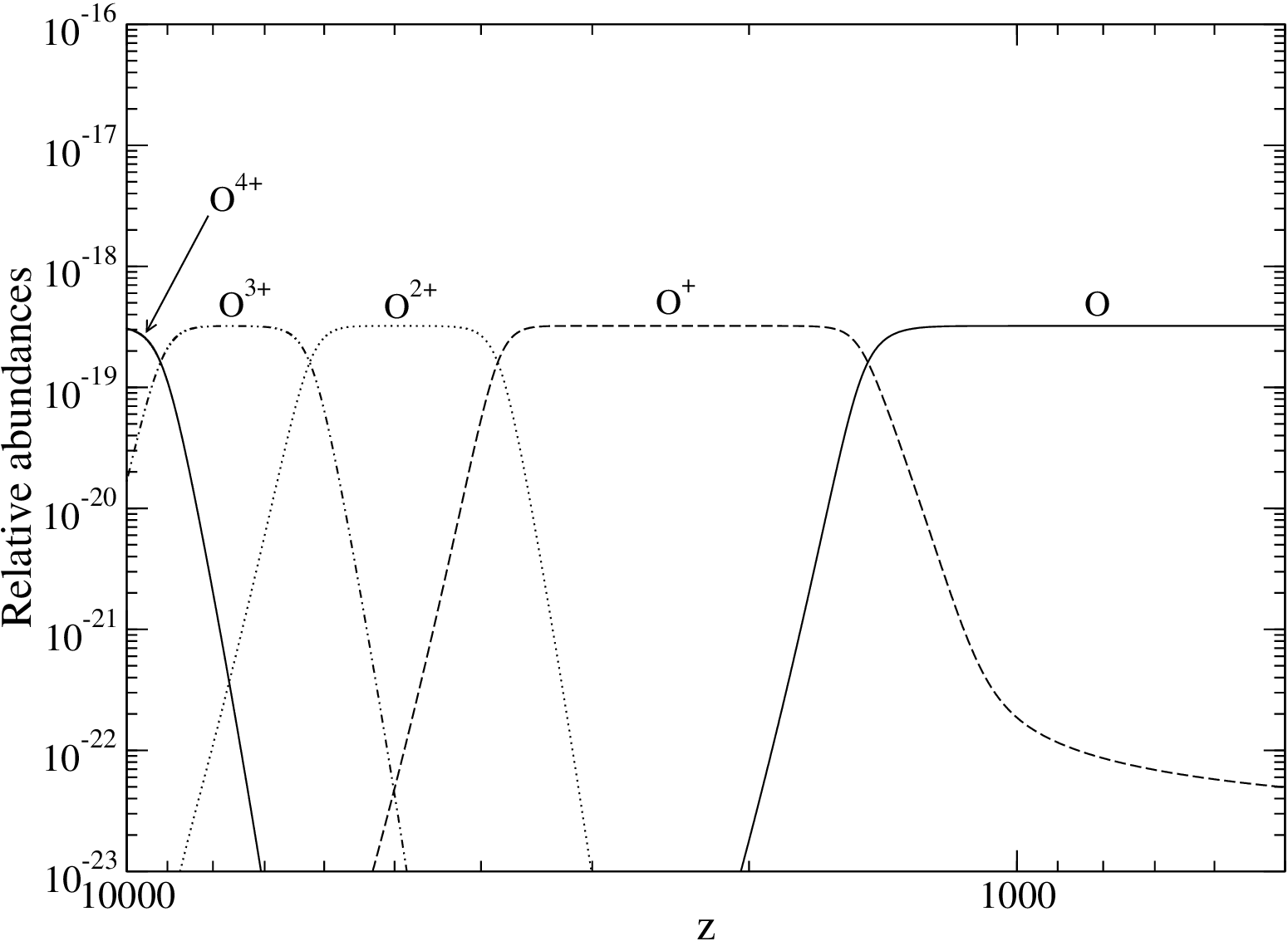}
\end{center}
\caption{Successive recombinations of the oxygen ions in the standard model.}
\label{fig:recombio}
\end{figure}

Moreover, we also investigated, for the first time, the primordial recombinations of the carbon, nitrogen and oxygen ions. The photoionization rates are calculated from the cross sections evaluated by \citet{ver96}. The recombination rates are taken from \citet{ver96a} for the CVI, CV, CIV, NVII, NVI, NV, OVIII, OVII and OVI recombinations. They come from the work of \citet{peq91} for the CIII, CII, NIV, NIII, NII and for all other oxygen recombinations. Finally, the recombination rates come from \citet{omu00} for the CI recombination and from the \textsc{umist} database\footnote{http://www.udfa.net} \citep{woodallumist} for the NI recombination. We plot in Figs. \ref{fig:recombic}, \ref{fig:recombin} and \ref{fig:recombio} the successive recombinations of carbon, nitrogen and oxygen ions in the SBBC model. Recombinations in the NSBBC scenarios are very close to the standard recombinations, the only difference being a shift in relative abundances corresponding to the difference in the initial conditions.

Since the ionization potential of carbon ($I_{\rm C}$ = 11.26 eV) is smaller than that of hydrogen ($I_{\rm H}$=13.6 eV), C$^+$ recombination occurs later than H$^+$ recombination. Figure \ref{fig:recombic} shows that carbon is singly ionized at the beginning of the primordial chemistry. We conclude that it is not necessary to compute all the successive recombinations when calculating the evolution of the carbon species. Thus, we will consider that all the carbon ions are present as C$^+$ at $z_{\mathrm{init}} = 10^4$. For similar arguments drawn from Figs. \ref{fig:recombin} and \ref{fig:recombio}, we will start our calculations of the primordial chemistry of nitrogen and oxygen with the singly ionized nitrogen and oxygen ions at $z_{\mathrm{init}}$.

\subsection{Molecular formation}
\label{heavySBBN}
After hydrogen recombination, the decrease of the electron density leads to the inefficiency of thermal coupling between matter and radiation. The chemical processes occuring during this post-recombination period are essentially collisional (ionizations, radiative recombinations, attachments, etc.) and radiative, due to the active presence of the CMB photons. The chemical network leads gradually to the formation of the first molecules in the Universe \citep{lep84,puy93,sta98,GP98,sig99,lep02,pfe03,puysig07,gloabel08,glosav08}.

In this section, we will consider the primordial chemistry of hydrogen, deuterium, helium, lithium, carbon, nitrogen, oxygen and fluorine in the SBBN and NSBBN contexts. We list in the Appendix the sets of reactions that we consider.

\subsubsection{Hydrogen}
The cosmological production of H$_2$ molecules proceeds mainly through two mechanisms which are catalyzed by H$^-$ and H$_2^+$ \citep{sas67,peedic68}. It is generally considered that the H$_2^+$ mechanism produces molecular hydrogen at $z \sim 300$, while the H$^-$ mechanism is responsible for the H$_2$ peak at $z \sim 100$.

The reaction rates we consider for hydrogen are taken from \citet{GP98}, except for the recombination and photoionization rates H1 and H2 \citep{abe97}, and for the reaction H4 ($\mathrm{H}^- + \gamma \rightarrow \mathrm{H} + \mathrm{e}^-$). For that reaction, we consider the important correction described in \citet{hir06} that takes into account the effects of the non-thermal radiation emitted during hydrogen recombination.

\begin{figure}
\centering
\vspace{0.3cm}
\includegraphics[scale=0.3]{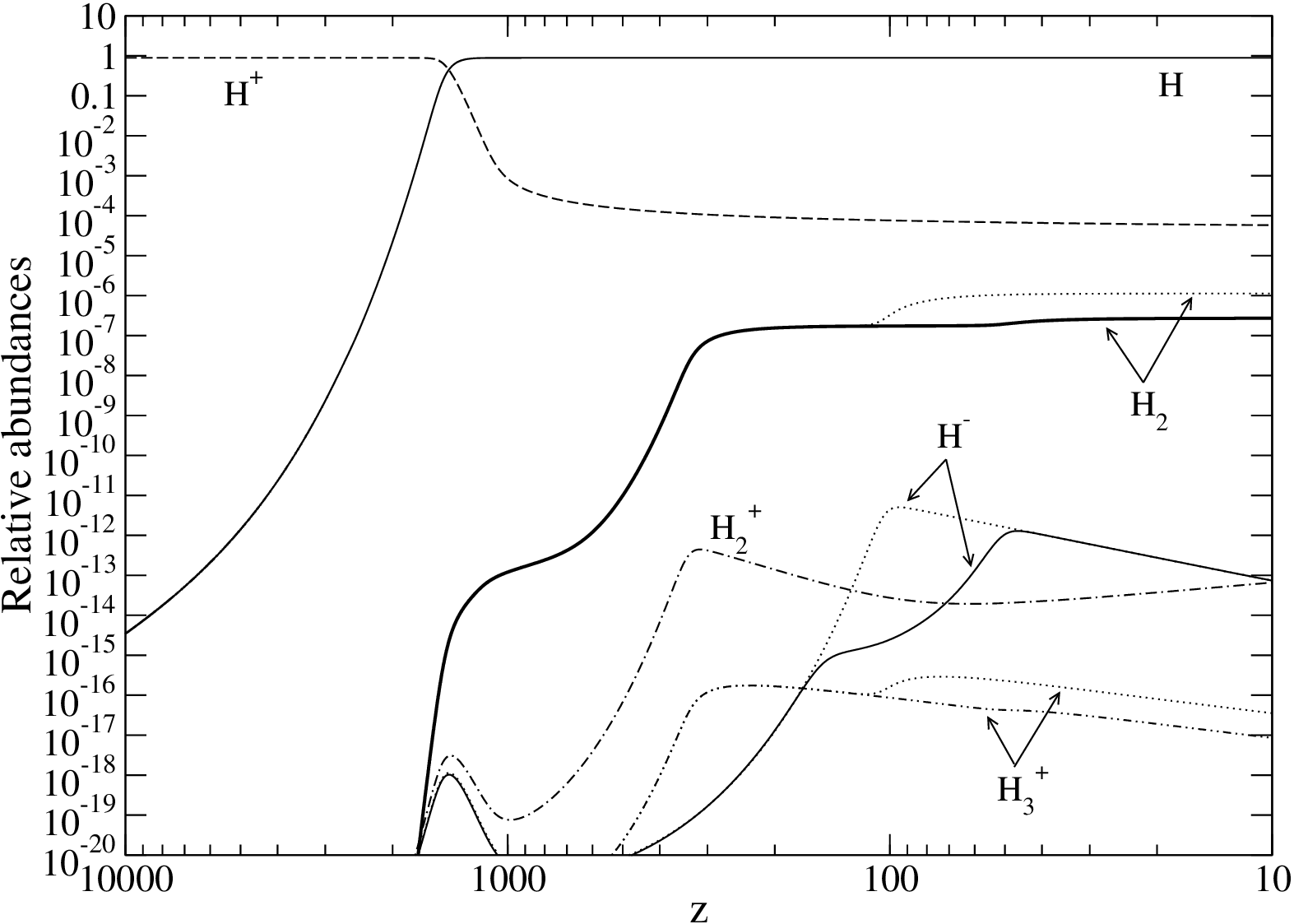}
\caption{Standard primordial chemistry of hydrogen species. The dotted lines for H$_2$, H$^-$ and H$_3^+$ show the abundances we would obtain without the correction of \citet{hir06} relative to the radiation emitted during the hydrogen recombination.}
\label{H}
\end{figure}

Figure \ref{H} shows the evolution of the hydrogen chemistry in the standard model. The amount of H$_2$ molecules created during the Dark Ages is important and converge to the following values at $z = 10$: [H$_2$]$_{\mathrm{SBBC}}$ = $n_{\mathrm{H_2}}/n_{\mathrm{b}}$ = 2.7 $\times 10^{-7}$. The effect of the correction of \citet{hir06} is shown on Fig. \ref{H}. The enhancement of the rate H4 is responsible for the efficient destruction of H$^-$ via photodetachment. As a consequence, the contribution of the H$^-$ channel to the final amount of molecular hydrogen is strongly reduced at $z \approx 100$ and the value of the standard final abundance is lower than the values usually seen in the literature (about a factor 4). Table \ref{tab:abundh} gives the abundances at redshifts $z=1000$, $z=100$ and $z=10$. The initial relative abundance of hydrogen is very close in our three chemistry models SBBC, NSBBC1 and NSBBC2 (see Table \ref{mfrac}), as is the case for deuterium, helium and to some extent lithium. For that reason we only show the results of the standard chemistry calculation for these four light elements.
\begin{table}[h!]
\caption{Relative abundances $n_{\xi} / n_{\mathrm{b}}$ of hydrogen species at redshifts $z=1000$, $z=100$ and $z=10$ in the SBBC model.}
\label{tab:abundh}
\centering
\begin{tabular}{l l l l l}
\hline
\hline
& & & SBBC &\\
\hline
Species && $z=1000$ & $z=100$ & $z=10$\\
\hline
$[{\rm H}]$ && 0.888 & 0.889 & 0.889\\
$[{\rm H}^+]$ && $8.5 \times 10^{-4}$ & $7.6 \times 10^{-5}$ & $5.8 \times 10^{-5}$\\
$[{\rm H}^-]$ && $5.7 \times 10^{-21}$ & $2.5 \times 10^{-15}$ & $7.2 \times 10^{-14}$\\
$[{\rm H_2}]$ && $1.2 \times 10^{-13}$ & $1.7 \times 10^{-7}$ & $2.7 \times 10^{-7}$\\
$[{\rm H_2}^+]$ &&7.7 $ \times 10^{-20}$ & $2.7 \times 10^{-14}$ & $6.7 \times 10^{-14}$\\
$[{\rm H_3}^+]$ &&3.0 $ \times 10^{-22}$ & $8.7 \times 10^{-17}$ & $8.4 \times 10^{-18}$\\
\hline
\end{tabular}
\end{table}

\subsubsection{Deuterium}
We take the reaction rates for deuterium chemistry from \citet{GP98}, except for the recombination and photoionization rates D1 and D2 \citep{abe97}, which are the same as H1 and H2 (see Appendix A). Figure \ref{D} and Table \ref{tab:abundd} show the evolution of the deuterium chemistry. The fact that the abundance of molecular hydrogen is reduced compared to the usual value of the literature has a direct implication on the HD quantity. Indeed, reactions D6 and D8 are the most important formation channels of HD. These reactions are collisions between neutral or charged D with H$_2$. As a consequence, the standard final amount of HD molecules is also lower (by a factor of 2) than the canonical value and we get [HD]$_{\mathrm{SBBC}}$ = $1.3 \times 10^{-10}$.
\begin{figure}
\vspace{0.3cm}
\centering
\includegraphics[scale=0.3]{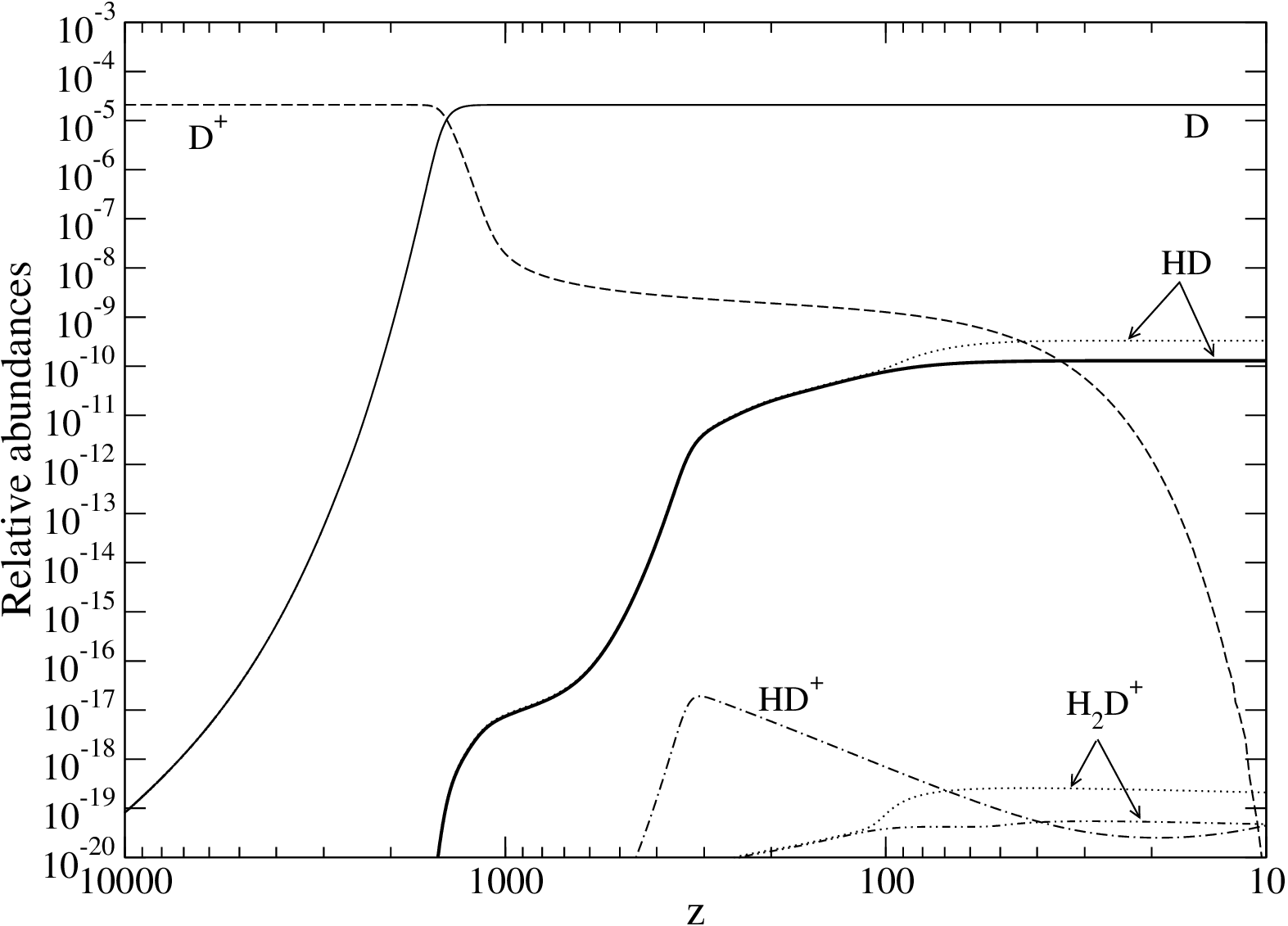}
\caption{Standard primordial chemistry of deuterium species. As in Fig. \ref{H}, the dotted lines for HD and H$_2$D$^+$ show the abundances without the correction of \citet{hir06} relative to the radiation emitted during the hydrogen recombination.}
\label{D}
\end{figure}
\begin{table}[h!]
\caption{Relative abundances $n_{\xi} / n_{\mathrm{b}}$ of deuterium species at redshifts $z=1000$, $z=100$ and $z=10$ in the SBBC model.}
\label{tab:abundd}
\centering
\begin{tabular}{l l l l l}
\hline
\hline
& & & SBBC &\\
\hline
Species && $z=1000$ & $z=100$ & $z=10$\\
\hline
$[{\rm D}]$ && $2.1 \times 10^{-5}$ & $2.1 \times 10^{-5}$ & $2.1 \times 10^{-5}$\\
$[{\rm D}^+]$ && $2.0 \times 10^{-8}$ & $1.3 \times 10^{-9}$ & $2.0 \times 10^{-21}$\\
$[{\rm HD}]$ && $7.4 \times 10^{-18}$ & $7.7 \times 10^{-11}$ & $1.3 \times 10^{-10}$\\
$[{\rm HD}^+]$ && $1.8 \times 10^{-24}$ & $6.9 \times 10^{-19}$ & $4.6 \times 10^{-20}$\\
$[{\rm H_2D}^+]$ && $2.1 \times 10^{-27}$ & $3.8 \times 10^{-20}$ & $4.7 \times 10^{-20}$\\
\hline
\end{tabular}
\end{table}

\subsubsection{Helium}
\begin{figure}
\vspace{0.3cm}
\centering
\includegraphics[scale=0.3]{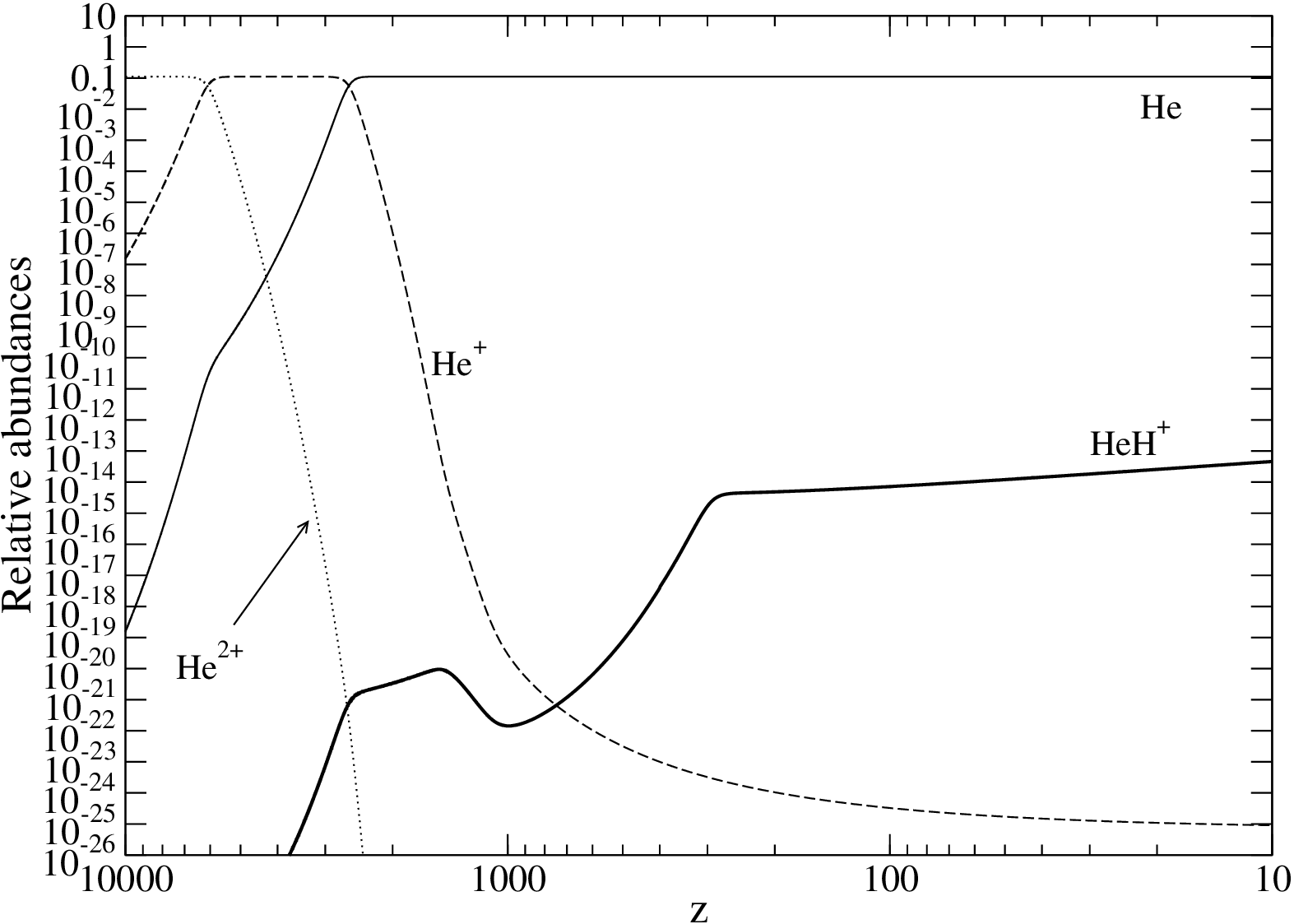}
\caption{Standard primordial chemistry of helium species.}
\label{He}
\end{figure}

The reaction rates we consider for helium are taken from \citet{GP98}. The results of the standard helium chemistry are shown in Fig. \ref{He}. In the first non-standard BBN model we consider (Sect. \ref{nsbbn}), the formation of helium nuclei during the primordial nucleosynthesis is extremely efficient in the neutron-rich, low-density regions. But primordial chemistry begins after the zones with different densities are mixed by baryon diffusion and the non-standard initial relative abundance of helium is identical to the standard one (see Table \ref{mfrac}). The main feature of helium chemistry is the production of molecule HeH$^+$ ([HeH$^+$]$_{\mathrm{SBBC}}$ = $4.6 \times 10^{-14}$ at $z = 10$). Table \ref{tab:abundhe} gives the abundances at redshifts $z=1000$, $z=100$ and $z=10$.
\begin{table}[h!]
\caption{Relative abundances $n_{\xi} / n_{\mathrm{b}}$ of helium species at redshifts $z=1000$, $z=100$ and $z=10$ in the SBBC model}
\label{tab:abundhe}
\centering
\begin{tabular}{l l l l l}
\hline
\hline
& & & SBBC &\\
\hline
Species && $z=1000$ & $z=100$ & $z=10$\\
\hline
$[{\rm He}]$ && 0.111 & 0.111 & 0.111\\
$[{\rm He}^+]$ && $3.0 \times 10^{-20}$ & $3.3 \times 10^{-25}$ & $9.1 \times 10^{-26}$\\
$[{\rm HeH}^+]$ && $1.4 \times 10^{-22}$ & $7.2 \times 10^{-15}$ & $4.6 \times 10^{-14}$\\
\hline
\end{tabular}
\end{table}

\subsubsection{Lithium}
\begin{figure}
\centering
\vspace{0.3cm}
\includegraphics[scale=0.3]{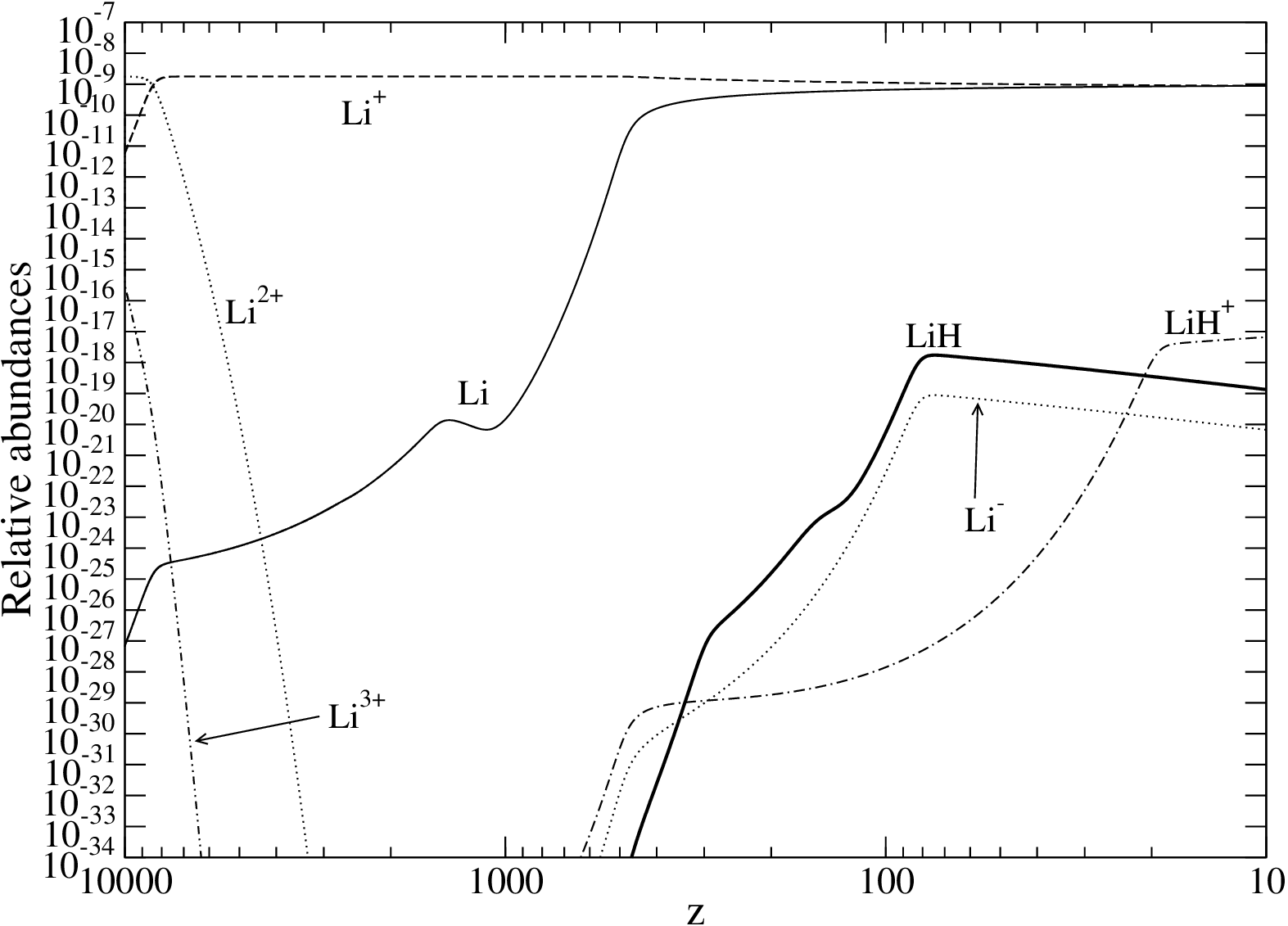}
\caption{Standard primordial chemistry of lithium species.}
\label{Li}
\end{figure}

We take the reaction rates for lithium chemistry from \citet{GP98}. Figure \ref{Li} and Table \ref{tab:abundli} show the evolution of the lithium species. As said before, lithium never recombines completely, and this is the reason why LiH is less abundant than LiH$^+$ at low redshift: [LiH]$_{\mathrm{SBBC}}$ = $1.3 \times 10^{-19}$ and [LiH$^+$]$_{\mathrm{SBBC}}$ = $6.7 \times 10^{-18}$ at $z = 10$.
\begin{table}[h!]
\caption{Relative abundances $n_{\xi} / n_{\mathrm{b}}$ of lithium species at redshifts $z=1000$, $z=100$ and $z=10$ in the SBBC model.}
\label{tab:abundli}
\centering
\begin{tabular}{l l l l l}
\hline
\hline
& & & SBBC &\\
\hline
Species && $z=1000$ & $z=100$ & $z=10$\\
\hline
$[{\rm Li}]$ && $1.4 \times 10^{-20}$ & $6.6 \times 10^{-10}$ & $8.8 \times 10^{-10}$\\
$[{\rm Li}^+]$ && $1.8 \times 10^{-9}$ & $1.1 \times 10^{-9}$ & $8.9 \times 10^{-10}$\\
$[{\rm Li}^-]$ && $4.4 \times 10^{-41}$ & $2.7 \times 10^{-22}$ & $6.6 \times 10^{-21}$\\
$[{\rm LiH}]$ && $9.5 \times 10^{-49}$ & $5.4 \times 10^{-21}$ & $1.3 \times 10^{-19}$\\
$[{\rm LiH}^+]$ && $1.5 \times 10^{-38}$ & $1.4 \times 10^{-28}$ & $6.7 \times 10^{-18}$\\
\hline
\end{tabular}
\end{table}

\subsubsection{Carbon}
The presence of neutral or ionized carbon at high redshift is important to determinate the cosmic background temperature at $z \neq 0$ in order to compare the result with models. \citet{son94,son95} used this technique to prove the existence of carbon in some high redshift diffuse gas. In the second paper, the detection of absorption due to the first level of neutral carbon fine structure, in a $z = 1.776$ cloud toward quasar Q1331+170, allows them to deduce that the background temperature is $7.4 \pm 0.8$ K at $z = 1.776$. This result agrees with the theoretical value (7.58 K). The same method was used by \citet{mol02}, who studied the fine structure levels of $\mathrm{C^+}$ in a Ly-$\alpha$ system at $z = 3.025$ toward Q0347-3819. Their result ($T_{\mathrm{CMB}} = 12.1^{+1.7}_{-3.2}$ K) agrees with the standard temperature limits.

A gas phase chemical model of the chemistry of CH and CH$^+$ was presented by \citet{dal76}, then \citet{pra80} presented a gas phase chemical model of the chemistry in interstellar clouds, including the C, N and O species. We consider here the evolution of the following carbon species: C, C$^+$, C$^-$, CH, CH$^+$, CH$_2$ and CH$_2^+$. The reaction rates are taken from the \textsc{umist} database \citep{woodallumist}, except for five reactions: reaction C7 \citep{lip07}, the two charge transfers C8 and C9 between C, H and their respective ions \citep{sta98a}, the recombination C48 \citep{omu00} and the photoionisation C49, whose cross section is taken from \citet{ver96} (see Table \ref{tab:reacc} of the Appendix). The main reaction leading to the synthesis of CH is clearly the radiative association C15 (C + H $\rightarrow$ CH + $\gamma$), but we also include the neutral reaction C12 and the two associative detachments C17 and C27. Among the numerous reactions destroying CH, the destructive collisional reactions with atomic hydrogen (reactions C10 and C11), and the charge exchange with H$^+$ (reaction C16) are the most important. The molecular ion CH$^+$ is created by the radiative association C$^+$ + H $\rightarrow$ CH$^+$ + $\gamma$ (reaction C23), the collisional reactions C14 and C18, and the charge exchanges C16 (with H$^+$) and C19 (with H$_2^+$), the main ones being C16 and C23. Destruction of CH$^+$ occurs mainly by collisions with H (reaction C25) and dissociative recombination (reaction C26).

In the reaction sets for carbon, nitrogen and oxygen chemistries (see Appendix A), we do not include photoionizations or photodissociations of primordial molecules (e. g. CH$^+$ + $\gamma$ $\rightarrow$ C$^+$ + H, NH$_2$ + $\gamma$ $\rightarrow$ NH$_2^+$ + e$^-$). These processes are especially important in the interstellar medium, where strong UV photons constitute the background radiation, whereas the Dark Ages molecules are embedded in a background radiation that is made of much softer microwave photons. In order to be sure not to neglect important photoprocesses, we made test calculations including photoionizations of the neutral molecules and photodissociations of all the molecules based on C, N and O. We computed the rates of these reactions using the CMB intensity and the constant cross section $\sigma = 10^{-17}$ cm$^2$. Indeed, this cross section gives a good order of magnitude estimate for the photoionization rates \citep{vanD88} and constitutes an upper limit for photodissociation processes. The results of the chemistry calculations were not changed when considering these photoprocesses.

\begin{figure}
\centering
\vspace{0.3cm}
\includegraphics[scale=0.3]{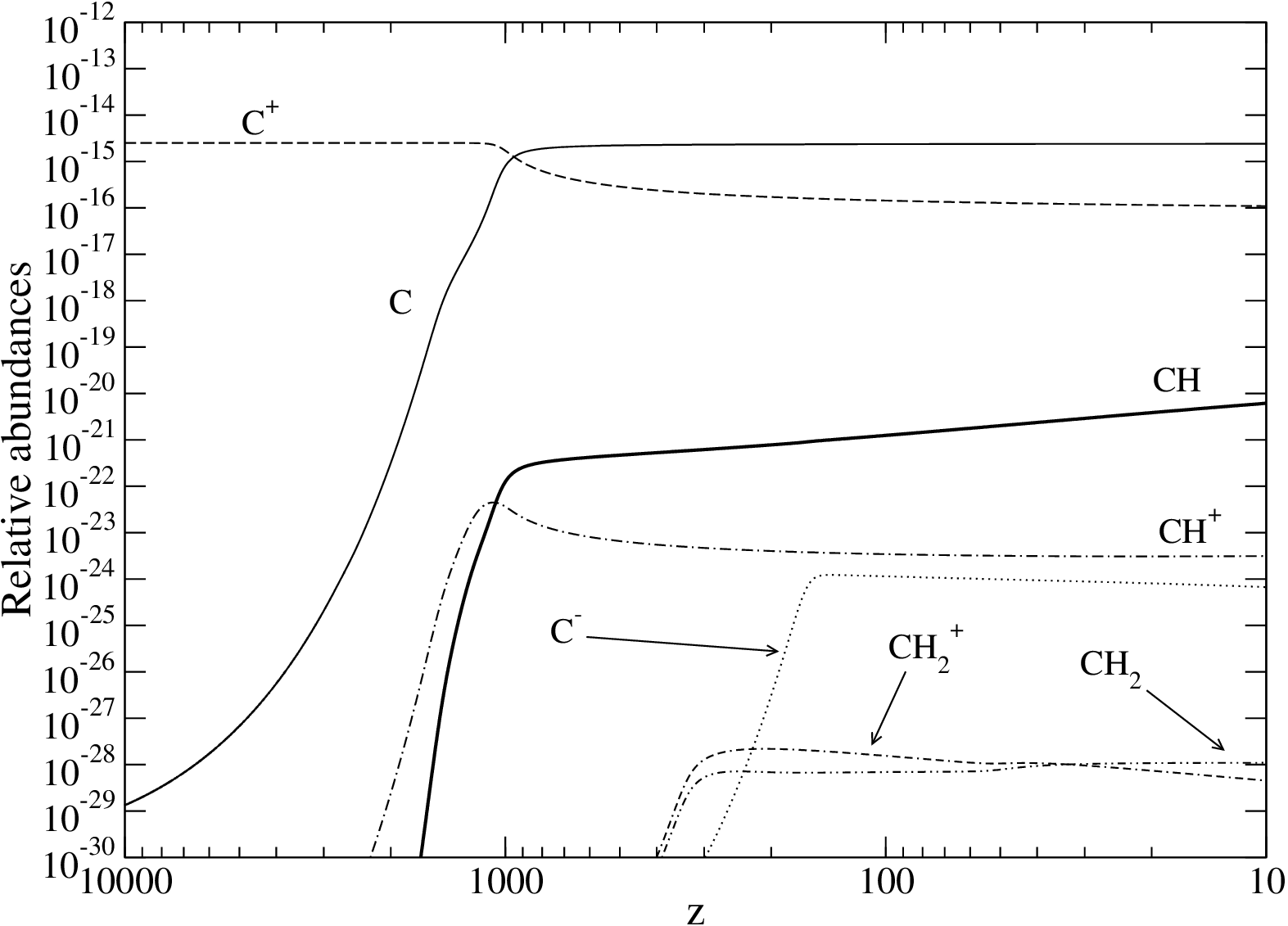}\\
\vspace{1cm}
\includegraphics[scale=0.3]{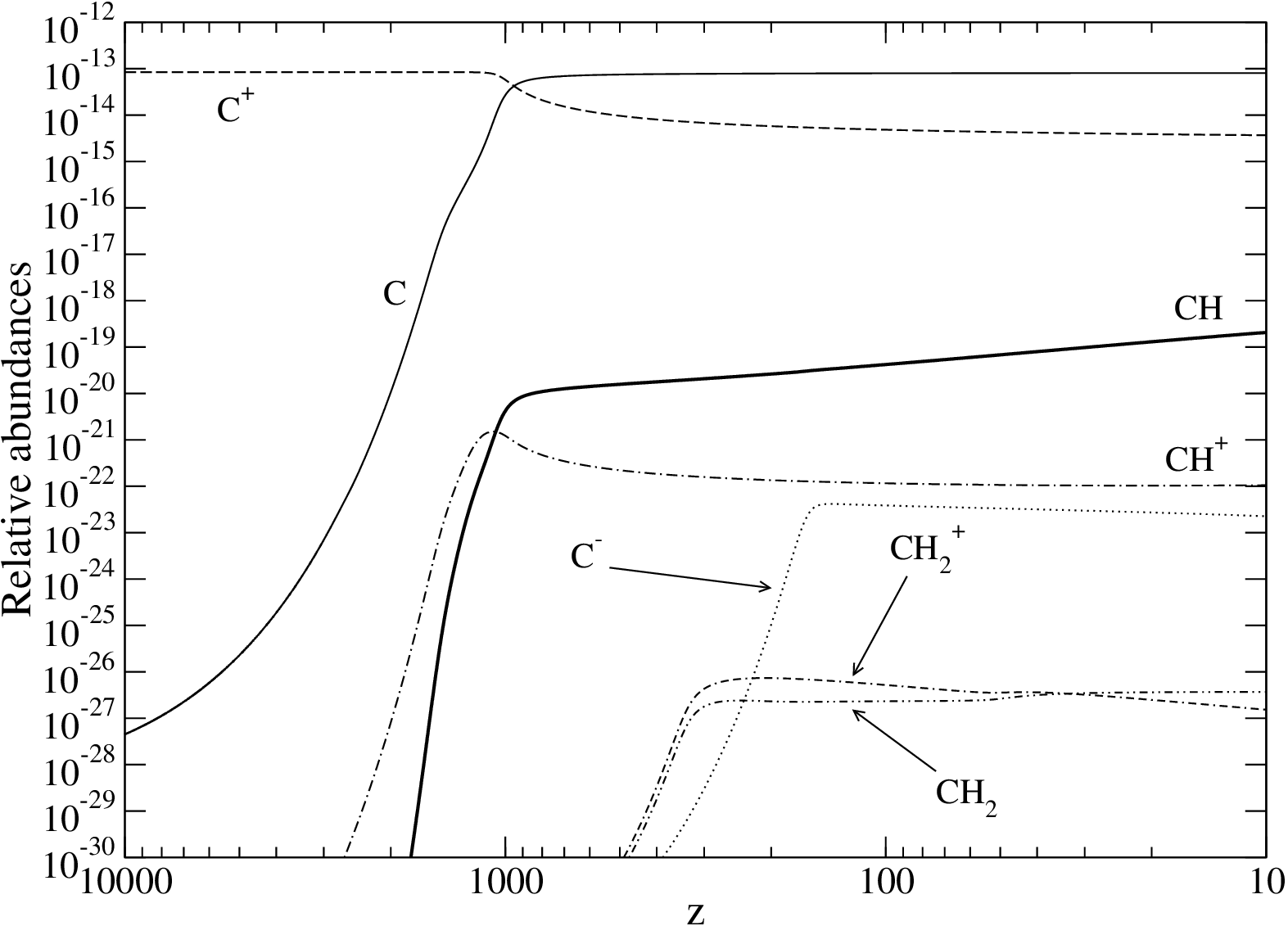}
\caption{Primordial chemistry of carbon species for the SBBC (upper panel) and NSBBC2 models (lower panel). The main difference between the two frames is a global increase of every relative abundance by a factor $\sim$ 30 in the non-standard model. Except for this shift, the evolution of the carbon species is similar in the two scenarios (see text). The results of the NSBBC1 model are quite close to the standard one.}
\label{C}
\end{figure}

Table \ref{tab:abundc} gives the abundances of the carbon species at redshifts $z=1000$, $z=100$ and $z=10$ for the three models SBBC, NSBBC1 and NSBBC2. Figure \ref{C} shows the evolution of these abundances as a function of redshift in the standard case and in the NSBBC2 case. We observe that the main difference between the standard and NSBBC2 runs is a global increase of every relative abundance by a factor $\sim$ 30 in the NSBBC2 run. Except for this shift, the evolution of every carbon species is quite similar in the two scenarios. This is not surprising, since the carbon chemistry is mainly determined by reactions with much more abundant hydrogen species like H, H$^+$, H$^-$, H$_2$ or H$_2^+$, and these species have the same abundances in all our models. The results of the NSBBC1 run are quite close to the standard one. We also note that the final amount of CH is about 20 times smaller than the LiH abundance in the standard model. This ratio is about 7.5 for HD$^+$ and H$_2$D$^+$. But we note that in the NSBBC2 case, molecule CH is as abundant as LiH and more abundant than HD$^+$ and H$_2$D$^+$ (about 4.5 times).

We note that the formation of carbon molecules is comparatively as efficient as the formation of H$_2$ or HD. Indeed, the ratio of the most abundant carbon molecule CH to C is $\mathrm{[CH]/[C]} \sim 2.6 \times 10^{-6}$ at $z = 10$ in our three models. For comparison, we have the following ratios fot the most abundant molecule based on each one of the light elements: $\mathrm{[H_2]/[H]} \sim 3 \times 10^{-7}$, $\mathrm{[HD]/[D]} \sim 6 \times 10^{-6}$, $\mathrm{[HeH^+]/[He]} \sim 4 \times 10^{-13}$ and $\mathrm{[LiH^+]/[Li]} \sim 8 \times 10^{-9}$ at $z = 10$.
\begin{table*}[ht!]
\caption{Relative abundances $n_{\xi} / n_{\mathrm{b}}$ of carbon species at redshifts $z=1000$, $z=100$ and $z=10$ in the SBBC (left), NSBBC1 (middle) and NSBBC2 (right) models.}
\label{tab:abundc}
\centering
\begin{tabular}{l l l l l l l l l l}
\hline
\hline
& & SBBC & & & NSBBC1 & & & NSBBC2 &\\
\hline
Species & $z=1000$ & $z=100$ & $z=10$ & $z=1000$ & $z=100$ & $z=10$ & $z=1000$ & $z=100$ & $z=10$\\
\hline
$[{\rm C}]$ & $7.9 \times 10^{-16}$ & $2.4 \times 10^{-15}$ & $2.4 \times 10^{-15}$ & $1.3 \times 10^{-15}$ & $3.8 \times 10^{-15}$ & $3.8 \times 10^{-15}$ & $2.7 \times 10^{-14}$ & $8.0 \times 10^{-14}$ & $8.1 \times 10^{-14}$\\
$[{\rm C}^+]$ & $1.7 \times 10^{-15}$ & $1.4 \times 10^{-16}$ & $1.1 \times 10^{-16}$ & $2.7 \times 10^{-15}$ & $2.3 \times 10^{-16}$ & $1.7 \times 10^{-16}$ & $5.7 \times 10^{-14}$ & $4.8 \times 10^{-15}$ & $3.7 \times 10^{-15}$\\
$[{\rm C}^-]$ & $7.6 \times 10^{-35}$ & $1.1 \times 10^{-24}$ & $6.7 \times 10^{-25}$ & $1.2 \times 10^{-34}$ & $1.8 \times 10^{-24}$ & $1.1 \times 10^{-24}$ & $2.6 \times 10^{-33}$ & $3.9 \times 10^{-23}$ & $2.3 \times 10^{-23}$\\
$[{\rm CH}]$ & $1.2 \times 10^{-22}$ & $1.2 \times 10^{-21}$ & $6.2 \times 10^{-21}$ & $2.0 \times 10^{-22}$ & $2.0 \times 10^{-21}$ & $9.8 \times 10^{-21}$ & $4.2 \times 10^{-21}$ & $4.2 \times 10^{-20}$ & $2.1 \times 10^{-19}$\\
$[{\rm CH}^+]$ & $3.7 \times 10^{-23}$ & $3.4 \times 10^{-24}$ & $3.1 \times 10^{-24}$ & $5.8 \times 10^{-23}$ & $5.5 \times 10^{-24}$ & $5.0 \times 10^{-24}$ & $1.2 \times 10^{-21}$ & $1.2 \times 10^{-22}$ & $1.1 \times 10^{-22}$\\
$[{\rm CH}_2]$ & $7.9 \times 10^{-35}$ & $6.9 \times 10^{-29}$ & $1.1 \times 10^{-28}$ & $1.3 \times 10^{-34}$ & $1.1 \times 10^{-28}$ & $1.7 \times 10^{-28}$ & $2.8 \times 10^{-33}$ & $2.3 \times 10^{-27}$ & $3.7 \times 10^{-27}$\\
$[{\rm CH}_2^+]$ & $2.7 \times 10^{-34}$ & $1.5 \times 10^{-28}$ & $4.5 \times 10^{-29}$ & $4.3 \times 10^{-34}$ & $2.5 \times 10^{-28}$ & $7.2 \times 10^{-29}$ & $9.2 \times 10^{-33}$ & $5.2 \times 10^{-27}$ & $1.5 \times 10^{-27}$\\
\hline
\end{tabular}
\end{table*}

\citet{lip07} investigated the formation of carbon molecules but with a higher initial ratio [C]/[H] = $10^{-10}$. They found a final CH abundance of $10^{-14}$, which means a similar efficiency toward molecular formation, even if their chemical model is different.

\subsubsection{Nitrogen}
We take the reaction rates for nitrogen chemistry from the \textsc{umist} database, except for the photoionisation N41 \citep{ver96} (see Table \ref{tab:reacn} of the Appendix) and study the evolution of N, N$^+$, NH, NH$^+$, NH$_2$ and NH$_2^+$. The main reactions responsible for the creation of NH are collisions of nitrogen atoms with H$_2$ and H$^-$ (reactions N5 and N14 respectively). We consider many processes that can lead to the destruction of NH, the most important of them being collisions with H (reaction N4) and the charge exchange H$^+$ + NH $\rightarrow$ NH$^+$ + H (reaction N10). The latter reaction is at the same time one of the main processes responsible for the synthesis of the molecular ion NH$^+$, together with the two collisional reactions N6 (between N and H$_2^+$) and N7 (between N$^+$ and molecular hydrogen). Among the differents processes destroying NH$^+$, the dissociative recombination N13 appears to be the most efficient.

\begin{figure}
\centering
\vspace{0.3cm}
\includegraphics[scale=0.3]{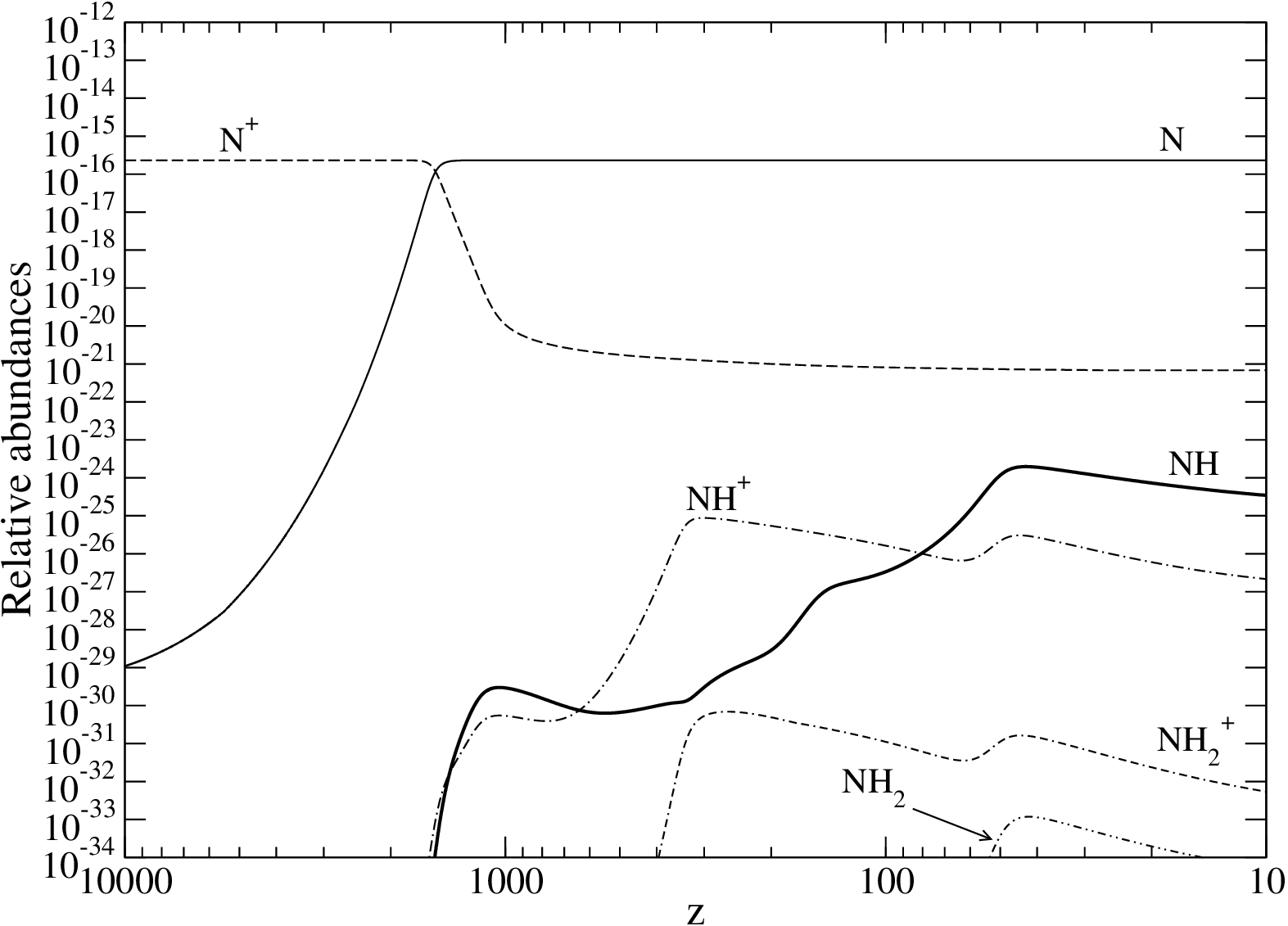}\\
\vspace{1cm}
\includegraphics[scale=0.3]{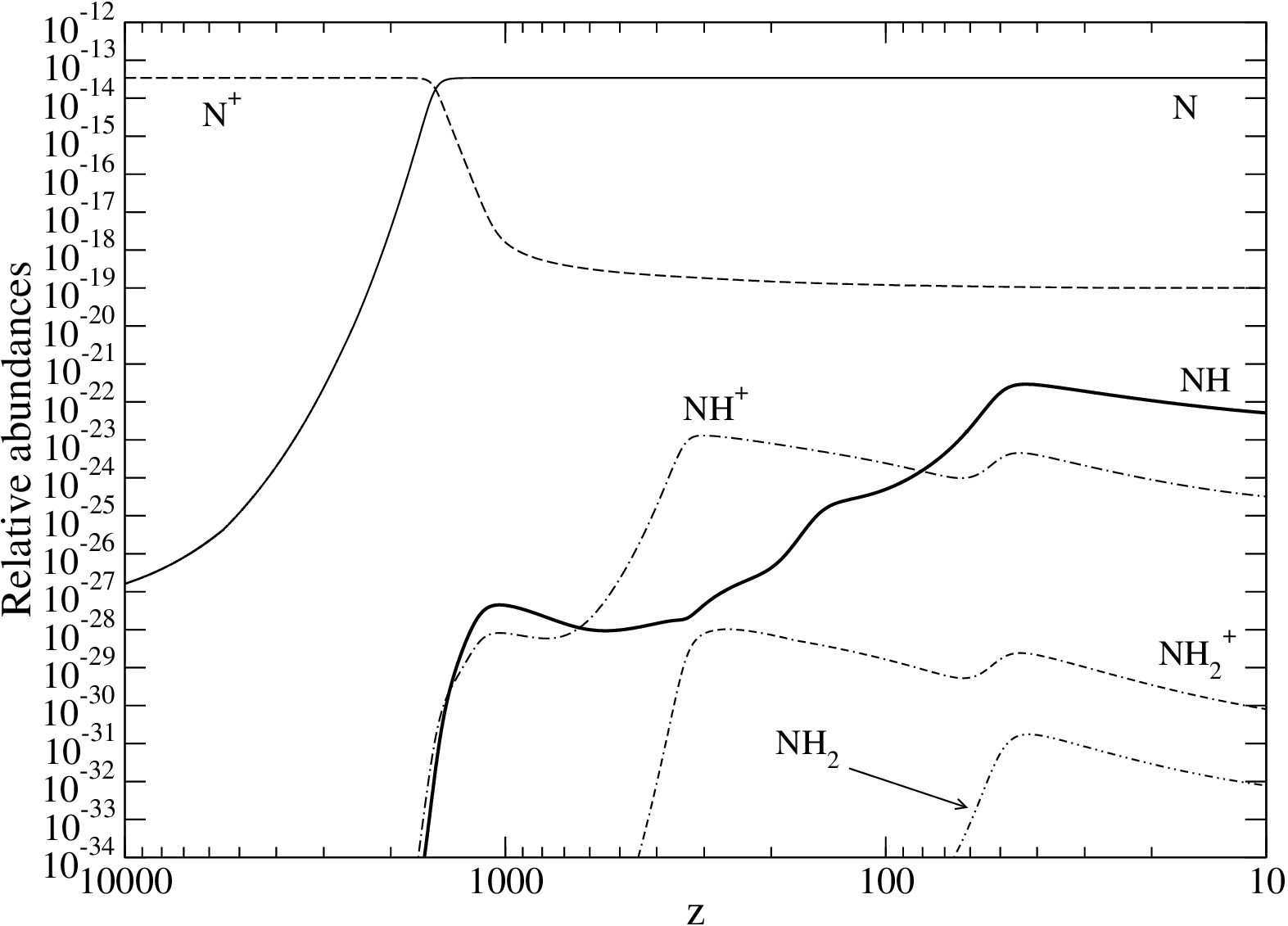}
\caption{Primordial chemistry of nitrogen species for the SBBC (upper panel) and NSBBC2 models (lower panel). The formation of molecules based on nitrogen is less effective than the formation of carbon molecules. The main difference between the two models shown here is a global increase of every relative abundance by more than two orders of magnitude in the non-standard model. Except for this shift, the evolution of the nitrogen species is similar in the two scenarios. The results of the NSBBC1 model are identical to the standard ones.}
\label{N}
\end{figure}

Figure \ref{N} shows the evolution of the nitrogen chemistry as a function of redshift in the standard and NSBBC2 cases. The molecular ion NH$^+$ dominates the nitrogen chemistry from $z \sim 600$ down to $z \sim 80$, and is then surpassed by NH, which has a peak around $z=40$. The final relative abundances at $z = 10$ are [NH]$_{\mathrm{SBBC}} = 3.4 \times 10^{-25}$ and [NH]$_{\mathrm{NSBBC2}} = 5.1 \times 10^{-23}$. As for carbon, the main difference between the SBBC and NSBBC2 scenarios is a global increase of every relative abundance (by a factor about 150) in the NSBBC2. Apart from this shift, the evolution of all the nitrogen species is quite similar in the two models. The results of the NSBBC1 run are almost identical to the standard one, since nitrogen nuclei are produced in the same amount in the two corresponding nucleosynthesis scenarios (Table \ref{mfrac}). Table \ref{tab:abundn} gives the abundances at redshifts $z=1000$, $z=100$ and $z=10$ for the three models.

The formation of molecules based on N is less effective than the formation of carbon molecules. When considering the most abundant species, NH, one has indeed: $\mathrm{[NH]/[N]} \sim 10^{-9}$ for the three scenarios.
\begin{table*}[ht!]
\caption{Relative abundances $n_{\xi} / n_{\mathrm{b}}$ of nitrogen species at redshifts $z=1000$, $z=100$ and $z=10$ in the SBBC (left), NSBBC1 (middle) and NSBBC2 (right) models.}
\label{tab:abundn}
\centering
\begin{tabular}{l l l l l l l l l l}
\hline
\hline
& & SBBC & & & NSBBC1 & & & NSBBC2 &\\
\hline
Species & $z=1000$ & $z=100$ & $z=10$ & $z=1000$ & $z=100$ & $z=10$ & $z=1000$ & $z=100$ & $z=10$\\
\hline
$[{\rm N}]$ & $2.3 \times 10^{-16}$ & $2.3 \times 10^{-16}$ & $2.3 \times 10^{-16}$ & $2.5 \times 10^{-16}$ & $2.5 \times 10^{-16}$ & $2.5 \times 10^{-16}$ & $3.4 \times 10^{-14}$ & $3.4 \times 10^{-14}$ & $3.4 \times 10^{-14}$\\
$[{\rm N}^+]$ & $1.1 \times 10^{-20}$ & $8.1 \times 10^{-22}$ & $6.8 \times 10^{-22}$ & $1.1 \times 10^{-20}$ & $8.6 \times 10^{-22}$ & $7.2 \times 10^{-22}$ & $1.6 \times 10^{-18}$ & $1.2 \times 10^{-19}$ & $1.0 \times 10^{-19}$\\
$[{\rm NH}]$ & $2.9 \times 10^{-30}$ & $3.4 \times 10^{-27}$ & $3.4 \times 10^{-25}$ & $3.1 \times 10^{-30}$ & $3.6 \times 10^{-27}$ & $3.7 \times 10^{-25}$ & $4.4 \times 10^{-28}$ & $5.0 \times 10^{-25}$ & $5.1 \times 10^{-23}$\\
$[{\rm NH}^+]$ & $5.4 \times 10^{-31}$ & $1.6 \times 10^{-26}$ & $2.1 \times 10^{-27}$ & $5.7 \times 10^{-31}$ & $1.7 \times 10^{-26}$ & $2.3 \times 10^{-27}$ & $8.1 \times 10^{-29}$ & $2.4 \times 10^{-24}$ & $3.2 \times 10^{-25}$\\
$[{\rm NH}_2]$ & $5.8 \times 10^{-44}$ & $3.4 \times 10^{-39}$ & $5.2 \times 10^{-35}$ & $6.2 \times 10^{-44}$ & $3.6 \times 10^{-39}$ & $5.6 \times 10^{-35}$ & $8.8 \times 10^{-42}$ & $5.1 \times 10^{-37}$ & $7.9 \times 10^{-33}$\\
$[{\rm NH}_2^+]$ & $2.1 \times 10^{-42}$ & $1.1 \times 10^{-31}$ & $5.3 \times 10^{-33}$ & $2.3 \times 10^{-42}$ & $1.2 \times 10^{-31}$ & $5.8 \times 10^{-33}$ & $3.3 \times 10^{-40}$ & $1.6 \times 10^{-29}$ & $8.1 \times 10^{-31}$\\
\hline
\end{tabular}
\end{table*}

\subsubsection{Oxygen}
\begin{figure}
\centering
\vspace{0.3cm}
\includegraphics[scale=0.3]{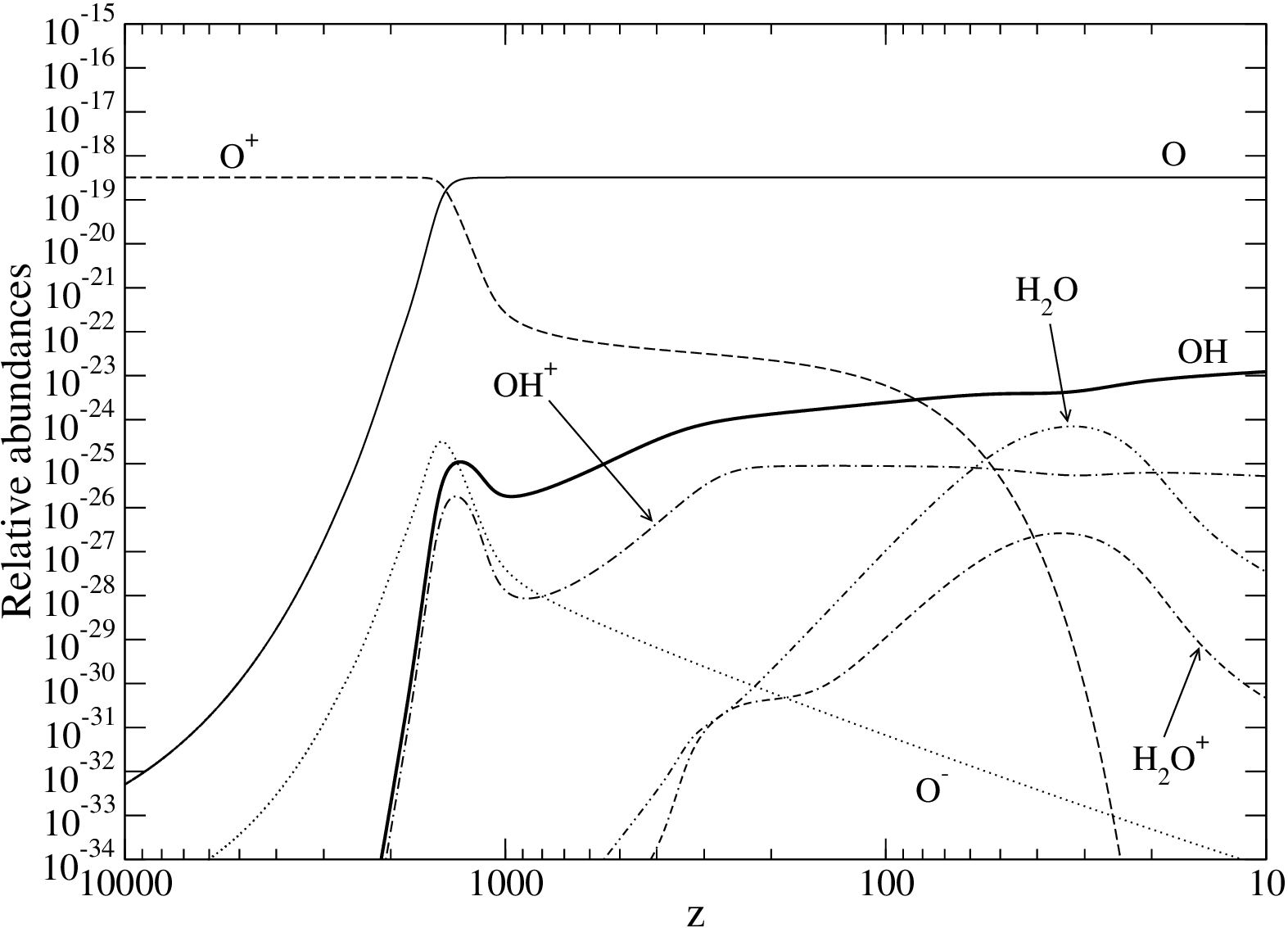}\\
\vspace{1cm}
\includegraphics[scale=0.3]{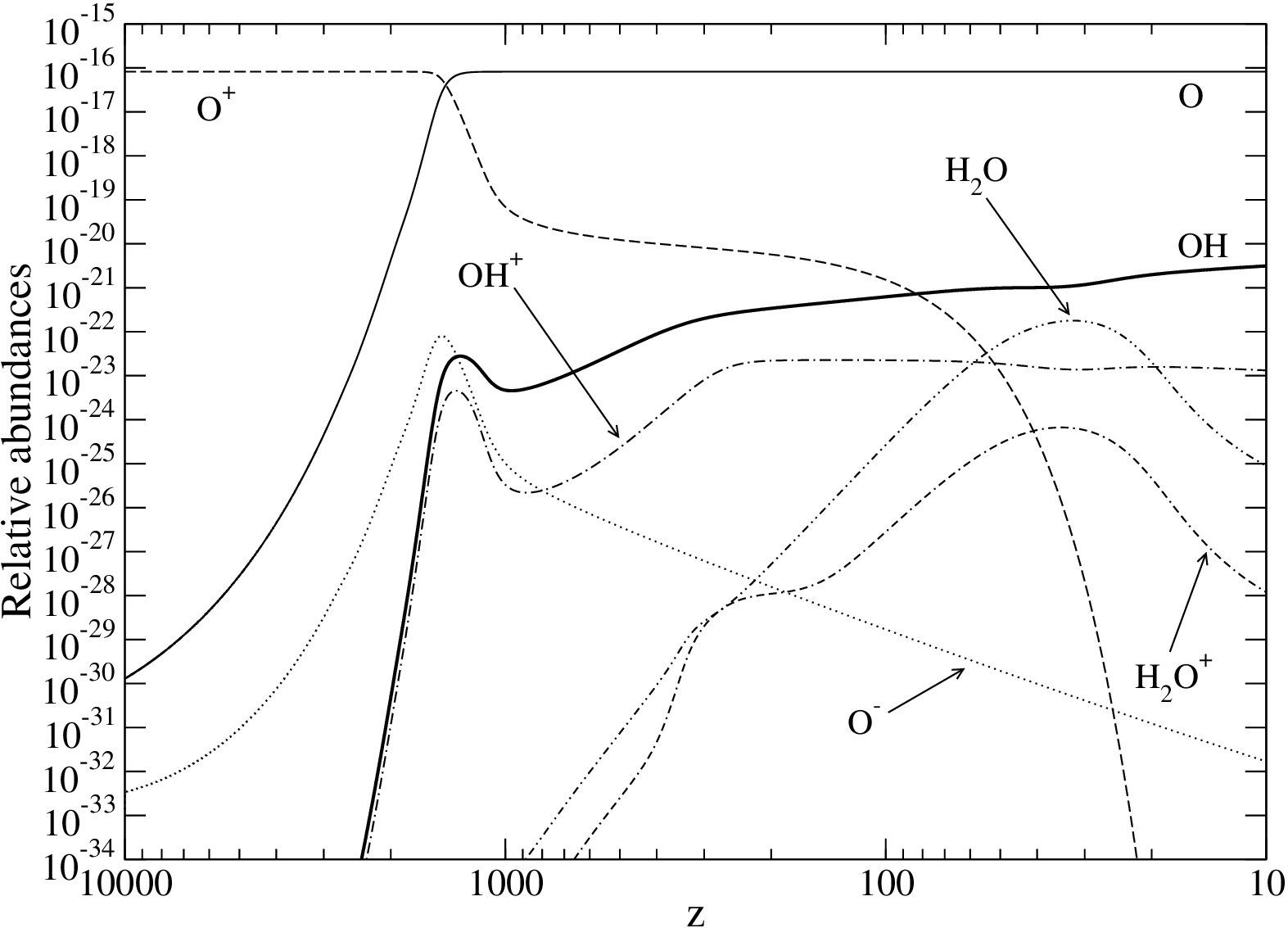}
\caption{Primordial chemistry of oxygen species for the SBBC (upper panel) and NSBBC2 models (lower panel). OH is clearly the most abundant oxygen molecule. Primordial water molecules show a peak around $z = 30$ but are then greatly depleted at lower redshifts. The main difference between the two models is a global increase of every relative abundance by more than two orders of magnitude in the non-standard NSBBC2 model. Except for this shift, the evolution of the oxygen species is similar in the two scenarios. The results of the NSBBC1 model are very close to the standard ones.}
\label{O}
\end{figure}

We consider the following species for oxygen chemistry: O, O$^+$, O$^-$, OH, OH$^+$, H$_2$O and H$_2$O$^+$. Table \ref{tab:reaco} shows the reactions we include in our chemical model. The rates for reactions O61 and O62 come from \citet{lebourlot1993}. The rate for recombination O58 is taken from \citet{peq91} and the cross section for photoionisation O59 from \citet{ver96}. All other reaction rates come from the \textsc{umist} database. The results of oxygen chemistry are shown in Fig. \ref{O} and Table \ref{tab:abundo} shows the relative abundances of the oxygen species at redshift $z=1000$, $z=100$ and $z=10$ for the three models. OH and OH$^+$ are the main molecular species based on oxygen at the end of the Dark Ages. In order to synthesize OH, we consider, among many other reactions, the radiative association O9, collisions between H and O$^-$ (reaction O10) and between H$^-$ and O (reaction O13), and the collisional reaction O15. Note that the first two processes are clearly the most effective. Numerous reactions destroy OH molecules, but collisions with H (reactions O11, O12 and O47) and H$^+$ (reaction O14) dominate. The molecular ion OH$^+$ is produced via the charge exchange O14 (H$^+$ + OH $\rightarrow$ OH$^+$ + H), the other reactions being less important. OH$^+$ is mainly destroyed by dissociative recombination (reaction O22). The collisions with hydrogen atoms (OH$^+$ + H $\rightarrow$ O$^+$ + H$_2$) are efficient at very high redshift only ($z > 300$). The very strong depletion of O$^-$ is caused by the photoprocess O$^-$ + $\gamma$ $\rightarrow$ O + e$^-$ (reaction O5). But the \textsc{umist} rate for that process assumes an interstellar UV background which obviously is inappropriate for our study. We thus expect a final O$^-$ abundance much higher than plotted in Fig. \ref{O}.

The tendency toward molecular formation is quite strong: we observe $\mathrm{[OH]/[O]} \sim 4 \times 10^{-5}$ at $z = 10$ in our three standard and non-standard calculations, and even if the initial oxygen abundances at $z = 10^4$ are very weak, OH is only one order of magnitude less abundant than HD$^+$ and H$_2$D$^+$ in the NSBBC2 run. We note the formation of primordial water molecules around $z = 30$. They are mainly synthesized by reaction O47 (H + OH $\rightarrow$ H$_2$O + $\gamma$) but at lower redshifts their abundance severly drops due to the destructive effect of reaction O60 (H$^+$ + H$_2$O $\rightarrow$ H$_2$O$^+$ + H). Again, as for carbon and nitrogen, the SBBC and NSBBC2 runs differ mainly by a global shift in relative abundance (by more than two orders of magnitude).
\begin{table*}[ht!]
\caption{Relative abundances $n_{\xi} / n_{\mathrm{b}}$ of oxygen species at redshifts $z=1000$, $z=100$ and $z=10$ in the SBBC (left), NSBBC1 (middle) and NSBBC2 (right) models.}
\label{tab:abundo}
\centering
\begin{tabular}{l l l l l l l l l l}
\hline
\hline
& & SBBC & & & NSBBC1 & & & NSBBC2 &\\
\hline
Species & $z=1000$ & $z=100$ & $z=10$ & $z=1000$ & $z=100$ & $z=10$ & $z=1000$ & $z=100$ & $z=10$\\
\hline
$[{\rm O}]$ & $3.2 \times 10^{-19}$ & $3.2 \times 10^{-19}$ & $3.2 \times 10^{-19}$ & $3.4 \times 10^{-19}$ & $3.4 \times 10^{-19}$ & $3.4 \times 10^{-19}$ & $8.2 \times 10^{-17}$ & $8.2 \times 10^{-17}$ & $8.2 \times 10^{-17}$\\
$[{\rm O}^+]$ & $2.7 \times 10^{-22}$ & $6.0 \times 10^{-24}$ & $0.0$ & $2.8 \times 10^{-22}$ & $6.2 \times 10^{-24}$ & $0.0$ & $6.9 \times 10^{-20}$ & $1.5 \times 10^{-21}$ & $0.0$\\
$[{\rm O}^-]$ & $4.1 \times 10^{-28}$ & $6.7 \times 10^{-32}$ & $6.4 \times 10^{-35}$ & $4.2 \times 10^{-28}$ & $7.0 \times 10^{-32}$ & $7.0 \times 10^{-35}$ & $1.0 \times 10^{-25}$ & $1.7 \times 10^{-29}$ & $1.7 \times 10^{-32}$\\
$[{\rm OH}]$ & $1.8 \times 10^{-26}$ & $2.5 \times 10^{-24}$ & $1.2 \times 10^{-23}$ & $1.9 \times 10^{-26}$ & $2.6 \times 10^{-24}$ & $1.3 \times 10^{-23}$ & $4.7 \times 10^{-24}$ & $6.2 \times 10^{-22}$ & $3.1 \times 10^{-21}$\\
$[{\rm OH}^+]$ & $1.3 \times 10^{-28}$ & $8.8 \times 10^{-26}$ & $5.2 \times 10^{-26}$ & $1.4 \times 10^{-28}$ & $9.2 \times 10^{-26}$ & $5.4 \times 10^{-26}$ & $3.3 \times 10^{-26}$ & $2.2 \times 10^{-23}$ & $1.3 \times 10^{-23}$\\
$[{\rm H_2O}]$ & $1.1 \times 10^{-37}$ & $1.0 \times 10^{-27}$ & $3.3 \times 10^{-28}$ & $1.1 \times 10^{-37}$ & $1.1 \times 10^{-27}$ & $3.7 \times 10^{-28}$ & $2.8 \times 10^{-35}$ & $2.7 \times 10^{-25}$ & $9.2 \times 10^{-26}$\\
$[{\rm H_2O}^+]$ & $2.9 \times 10^{-39}$ & $1.1 \times 10^{-29}$ & $4.4 \times 10^{-31}$ & $3.1 \times 10^{-39}$ & $1.2 \times 10^{-29}$ & $4.9 \times 10^{-31}$ & $7.6 \times 10^{-37}$ & $2.8 \times 10^{-27}$ & $1.2 \times 10^{-28}$\\
\hline
\end{tabular}
\end{table*}

\subsubsection{Fluorine}
Even if the abundance of primary fluorine is much below the abundances of primordial C, N and O at the end of the Big Bang nucleosynthesis, the molecule HF could be efficiently formed in the early Universe for two reasons: the ionization potential of F is greater than that of hydrogen, and the binding energy of the hydrogen atom in HF is greater than the binding energy of H$_2$. The chemistry of fluorine in the post-recombination epoch has been studied in \citet{puy07}. We use here the same set of reactions (see Table \ref{tab:reacf} of the Appendix).

Figure \ref{F} and Table \ref{tab:abundf} show the evolution of the fluorine chemistry as a function of redshift in the standard and NSBBC2 cases. We clearly see that the fluorine chemistry is comparatively more efficient than the CNO chemistries: $\mathrm{[HF]/[F]} \sim 10^{-2}$ in the three models. The enhancement factor for HF in the NSBBC2 model compared to the standard value is 500.
\begin{figure}
\centering
\vspace{0.3cm}
\includegraphics[scale=0.3]{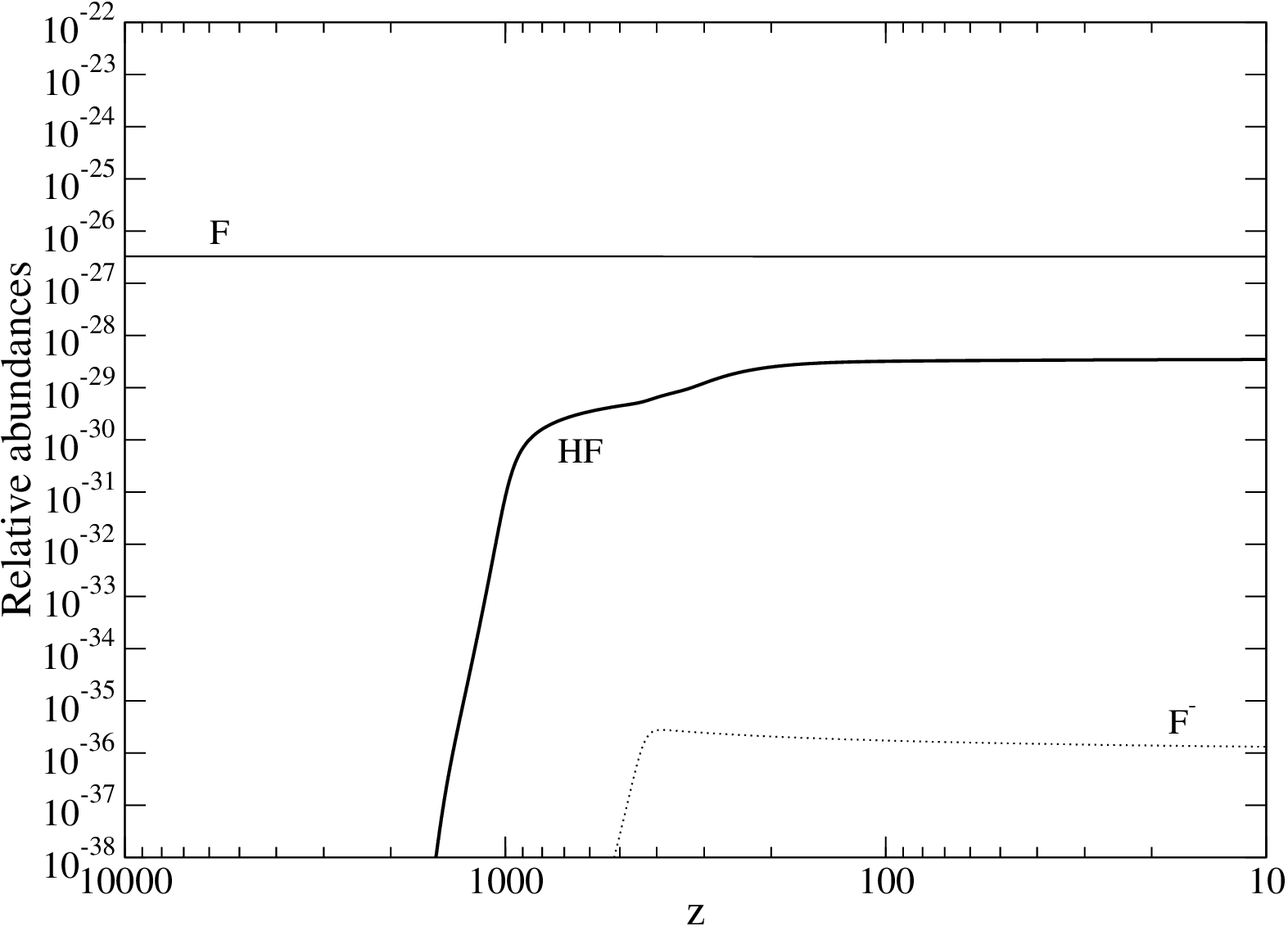}\\
\vspace{1cm}
\includegraphics[scale=0.3]{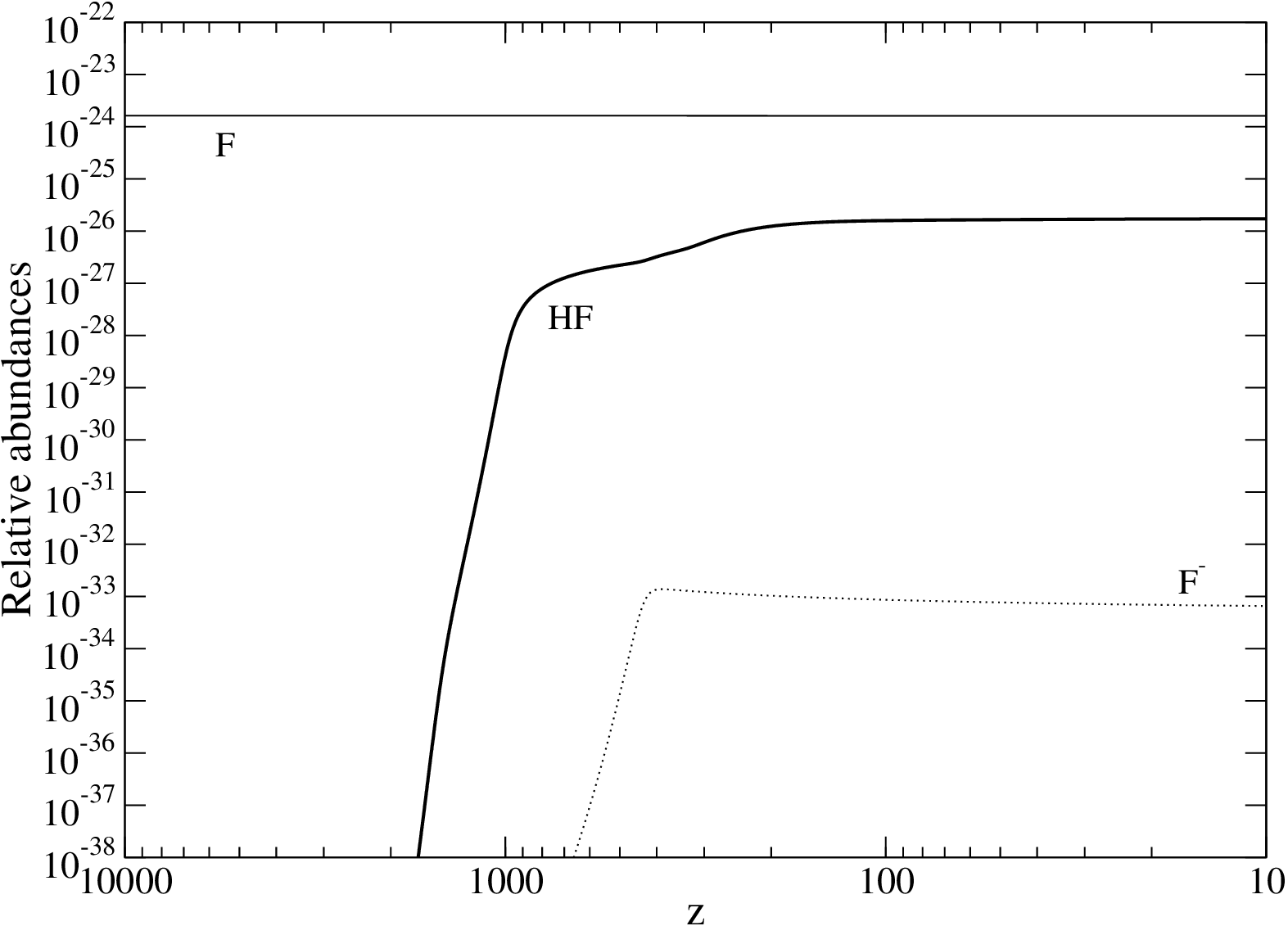}
\caption{Primordial chemistry of fluorine species for the SBBC (upper panel) and NSBBC2 models (lower panel). HF is more efficiently created by a factor 500 in the NSBBC2. The results of the NSBBC1 model are almost identical to the standard ones.}
\label{F}
\end{figure}
\begin{table*}[ht!]
\caption{Relative abundances $n_{\xi} / n_{\mathrm{b}}$ of fluorine species at redshifts $z=1000$, $z=100$ and $z=10$ in the SBBC (left), NSBBC1 (middle) and NSBBC2 (right) models.}
\label{tab:abundf}
\centering
\begin{tabular}{l l l l l l l l l l}
\hline
\hline
& & SBBC & & & NSBBC1 & & & NSBBC2 &\\
\hline
Species & $z=1000$ & $z=100$ & $z=10$ & $z=1000$ & $z=100$ & $z=10$ & $z=1000$ & $z=100$ & $z=10$\\
\hline
$[{\rm F}]$ & $3.3 \times 10^{-27}$ & $3.2 \times 10^{-27}$ & $3.2 \times 10^{-27}$ & $3.6 \times 10^{-27}$ & $3.6 \times 10^{-27}$ & $3.6 \times 10^{-27}$ & $1.6 \times 10^{-24}$ & $1.6 \times 10^{-24}$ & $1.6 \times 10^{-24}$\\
$[{\rm F}^-]$ & $2.6 \times 10^{-44}$ & $1.7 \times 10^{-36}$ & $1.3 \times 10^{-36}$ & $2.9 \times 10^{-44}$ & $1.9 \times 10^{-36}$ & $1.5 \times 10^{-36}$ & $1.3 \times 10^{-41}$ & $8.6 \times 10^{-34}$ & $6.6 \times 10^{-34}$\\
$[{\rm HF}]$ & $7.8 \times 10^{-32}$ & $3.2 \times 10^{-29}$ & $3.5 \times 10^{-29}$ & $9.1 \times 10^{-32}$ & $3.5 \times 10^{-29}$ & $3.8 \times 10^{-29}$ & $4.1 \times 10^{-29}$ & $1.6 \times 10^{-26}$ & $1.7 \times 10^{-26}$\\
\hline
\end{tabular}
\end{table*}

\section{Summary and Conclusion}
\label{sect:concl}
Molecules play an important role in the process of gravitational collapse through the mechanism of thermal instability \citep[see][]{field65,silk77,fall85,ueh96,puy96,abel00}. They provide an important cooling mechanism for primordial metal-free gas in different primordial contexts such as small haloes virializing at high redshift \citep[e.g.][]{bar01,bromm04,ciardi05} or early structure formation \citep{glo07,maio07}.

The H$_2$ molecule has long been recognized as the most important cooling agent, despite the absence of dipole moment, due to its high relative abundance in comparison with other molecules. For this reason H$_2$ molecules are a key component of many dynamical situations and particularly in the context of a gravitational collapse. However, several studies \citep{teg97,omu98,abel00,rip02,yosh06} consider H$_2$ as the only molecular coolant of the primordial gas, although HD can play an important role at low temperature. Indeed, \citet{puy97} showed that, if the primordial gas cools below 200 K, HD molecules are the main cooling agent, despite their low number abundance. These results were confirmed in different astrophysical mediums \citep{flo00,ueh00,nak00,flo01,lip05,nag05,sh06,rip07}. Recently, \citet{pri08} analyzed the effects of H$_2$, HD and LiH molecules in the cooling of primordial gas. Their simulations clearly showed that the gas, at low densities, reaches temperatures about 100 K and that the main coolant is H$_2$. But at higher densities ($n > 10^2$ cm$^{-3}$) HD dominates and the gas cools well below 100 K. The effects of LiH are negligible in all cases.

We analyzed in this paper the Dark Ages chemistry of eight elements (hydrogen, deuterium, helium, lithium, carbon, nitrogen, oxygen and fluorine), with the aim of determining the amount of primordial molecules based on elements heavier than lithium. Heavier molecules with high electric dipole moments could be important, if created in sufficient quantities, during the formation of the first structures in the Universe. We considered initial elemental abundances taken from two different nucleosynthesis contexts: the standard Big Bang nucleosynthesis model and non-standard nucleosynthesis based on baryon density fluctuations. In both contexts, the baryon-to-photon ratio $\eta$ (the averaged value for the non-standard model) was in accordance with the latest result of the WMAP experiment ($\eta_{10}^\mathrm{WMAP} = 6.225 \pm 0.170$), and the resulting abundances agreed with the observed primordial abundances of deuterium and helium.



The most abundant Dark Ages molecules based on elements heavier than lithium are CH and OH. Chemistry calculation assuming standard BBN initial abundances yields $\mathrm{[CH]_{SBBC}} = 6.2 \times 10^{-21}$ and $\mathrm{[OH]_{SBBC}} = 1.2 \times 10^{-23}$ at $z = 10$. NH molecules are less abundant: $\mathrm{[NH]_{SBBC}} = 3.4 \times 10^{-25}$.

We also made two non-standard chemistry calculations, considering two specific cases of inhomogenous BBN. In the first one, primordial nucleosynthesis abundances for C, N, O and F are only very slightly higher than in the standard case. For that reason, the chemistry of elements heavier than lithium in that context is very similar to the standard chemistry. But in the second non-standard BBN case, the final heavy element abundances are enhanced by globally two orders of magnitude. In that case, the relative abundances of molecular species based on carbon, nitrogen, oxygen or fluorine are modified by essentially the same factor, and molecules CH are as abundant as LiH and more abundant than HD$^+$ and H$_2$D$^+$: $\mathrm{[CH]_{NSBBC2}} = 2.1 \times 10^{-19}$ at $z = 10$. The details of the chemistry (e. g. the most important species, the most effective reactions, \ldots) are weakly affected.

\citet{gloabel08} investigated the effects of the chemical rate coefficient uncertainties in the Dark Ages chemistry of light elements. They showed that the large uncertainties in the associative detachment and mutual neutralization rates have an impact on the thermal evolution of the gas. As regards the heavy molecules discussed in the present work, the main reactions creating or destroying them are affected by uncertainties that can come to a factor 2. Moreover, some of the rates included in our model are given for a more restricted temperature range than the one of our calculations (from $z = 10^4$ to $z = 10$ the Universe cools from $T \sim 30000$ K down to a few Kelvins). We nonetheless assumed that these rates were valid everywhere in that range. Concerning oxygen chemistry, the \textsc{UMIST} rate for reaction O5 (O$^- + \gamma$ $\rightarrow$ O + e$^-$) is largely not appropriate to the Dark Ages chemistry. As one of the two main reactions creating OH is H + O$^-$ $\rightarrow$ OH + e$^-$, the use of a better rate for O5 could increase the OH relative abundance obtained in this work.

Knowledge of primordial molecular abundances of heavy molecules could be important in different contexts, cosmologically as well as for the formation of the first structures and stars. A cosmological influence of primordial molecules was suggested by \citet{dub77}. The measure of CMB anisotropies gives accurate indications on the fluctuation spectrum which led to present great structures. In this way, it is possible that primordial molecular clouds affect the anisotropy spectrum at small scales via resonant scattering of CMB photons on primordial molecules \citep[][for a review see \citealt{basu07} or \citealt{sig08}]{maoli94,maoli96,schleicher08}. The first spectral line surveys searching for primordial signals have been done with the Odin satellite \citep{persson09}. Their work may be considered as a pilot study for the forthcoming observations with the Herschel Space Observatory launched on May 14, 2009. One of the most promising molecule has been proven to be HeH$^+$. Molecules such as CH, NH or OH have electric dipole moments of the same order as HeH$^+$. We have shown in this work that their abundances are smaller by several orders of magnitude than HeH$^+$ abundance. But the fact that CH could be more abundant than molecules such as HD$^+$ and H$_2$D$^+$ could open new interesting perspectives and detailed calculations have still to be done.
In addition to molecules, heavy atoms and ions can also be used as a tool to probe the early Universe. \citet{bas04} considered the opacity generated by the scattering of CMB photons on heavy atoms like carbon, oxygen, silicon or iron (and on their respective ions). They found that the Planck HFI detectors will be able to get strong constraints ($10^{-3} - 10^{-4}$ solar abundance) on the most important atoms and ions (CNO) in the interval $z \in [5,30]$. \citet{har03} also examined the possibility to detect the imprint left by the presence of carbon or oxygen on the CMB during the Dark Ages.


Numerical simulations of star formation are a very important and popular problem in modern cosmology. If solved successfully, it can add a lot of information to our knowledge of the evolution of all kinds of cosmic structures and the Universe as a whole. Star formation is determined by complex chemical, dynamical and thermodynamical processes \citep[see][]{mcg08}, each of which plays an important role during some or all parts of the whole process, and the impact of cooling by heavier molecules on the formation of the first (Population III) stars is a new open question. Moreover, it is interesting to note that these first stars were themselves an important source of metallicity \citep{cooke09}, and cooling by heavier molecules could be even more important for later star formation.

\begin{acknowledgements}
We acknowledge the PNCG (Programme National Cosmologie et Galaxies) and the PNPS (Programme National de Physique Stellaire) for their financial assistance, and Daniel Pfenniger, Anton Lipovka and Dahbia Talbi for fruitful discussions. This work was in part supported by the Swiss National Science Foundation (grant 200020-122287).
\end{acknowledgements}

\bibliographystyle{aa} 
\bibliography{chemheavybib}

\begin{appendix}
\section{Reaction rates for the post-recombination chemistry}
\label{app}
We list in Tables \ref{tab:reach} to \ref{tab:reacf} the chemical reactions considered in our calculations.

\begin{table}[!h]
\caption{List of reactions for the hydrogen chemistry. The rates are taken from \citet{GP98}, except for the recombination and photoionization rates H1 and H2 \citep{abe97}, and for reaction H4 \citep{hir06}.}
\label{tab:reach}
\centering
\scriptsize
\begin{tabular}{l l l l}
\hline
\hline
& Reaction & & Reaction\\
\hline
H1 & $\mathrm{H^+ + e^- \rightarrow H + \gamma}$ & H2 & $\mathrm{H + \gamma \rightarrow H^+ + e^-}$\\
H3 & $\mathrm{H + e^- \rightarrow H^- + \gamma}$ & H4 & $\mathrm{H^- + \gamma \rightarrow H + e^-}$\\
H5 & $\mathrm{H^- + H \rightarrow H_2 + e^-}$ & H6 & $\mathrm{H^- + H^+ \rightarrow H_2^+ + e^-}$\\
H7 & $\mathrm{H^- + H^+ \rightarrow H + H}$ & H8 & $\mathrm{H + H^+ \rightarrow H_2^+ + \gamma}$\\
H9 & $\mathrm{H_2^+ + \gamma \rightarrow H + H^+}$ & H10 & $\mathrm{H_2^+ + H \rightarrow H_2 + H^+}$\\
H11 & $\mathrm{H_2^+ + e^- \rightarrow H + H}$ & H12 & $\mathrm{H_2^+ + \gamma \rightarrow H^+ + H^+ + e^-}$\\
H13 & $\mathrm{H_2^+ + H_2 \rightarrow H_3^+ + H}$ & H15 & $\mathrm{H_2 + H^+ \rightarrow H_2^+ + H}$\\
H16 & $\mathrm{H_2 + e^- \rightarrow H + H^-}$ & H17 & $\mathrm{H_2 + e^- \rightarrow H + H + e^-}$\\
H18 & $\mathrm{H_2 + \gamma \rightarrow H_2^+ + e^-}$ & H19 & $\mathrm{H_3^+ + H \rightarrow H_2^+ + H_2}$\\
H20 & $\mathrm{H_3^+ + e^- \rightarrow H_2 + H}$ & H21 & $\mathrm{H_2 + H^+ \rightarrow H_3^+ + \gamma}$\\
\hline
\end{tabular}
\end{table}

\begin{table}[!h]
\caption{List of reactions for the deuterium chemistry. We take the rates from \citet{GP98}, except for the recombination and photoionization rates D1 and D2, which are the same as H1 and H2 \citep{abe97}.}
\label{tab:reacd}
\centering
\scriptsize
\begin{tabular}{l l l l}
\hline
\hline
& Reaction & & Reaction\\
\hline
D1 & $\mathrm{D^+ + e^- \rightarrow D + \gamma}$ & D2 & $\mathrm{D + \gamma \rightarrow D^+ + e^-}$\\
D3 & $\mathrm{D + H^+ \rightarrow D^+ + H}$ & D4 & $\mathrm{D^+ + H \rightarrow D + H^+}$\\
D5 & $\mathrm{D + H \rightarrow HD + \gamma}$ & D6 & $\mathrm{D + H_2 \rightarrow H + HD}$\\
D7 & $\mathrm{HD^+ + H \rightarrow H^+ + HD}$ & D8 & $\mathrm{D^+ + H_2 \rightarrow H^+ + HD}$\\
D9 & $\mathrm{HD + H \rightarrow H_2 + D}$ & D10 & $\mathrm{HD + H^+ \rightarrow H_2 + D^+}$\\
D11 & $\mathrm{HD + H_3^+ \rightarrow H_2 + H_2D^+}$ & D12 & $\mathrm{D + H^+ \rightarrow HD^+ + \gamma}$\\
D13 & $\mathrm{D^+ + H \rightarrow HD^+ + \gamma}$ & D14 & $\mathrm{HD^+ + \gamma \rightarrow H + D^+}$\\
D15 & $\mathrm{HD^+ + \gamma \rightarrow H^+ + D}$ & D16 & $\mathrm{HD^+ + e^- \rightarrow H + D}$\\
D17 & $\mathrm{HD^+ + H_2 \rightarrow H_2D^+ + H}$ & D18 & $\mathrm{HD^+ + H_2 \rightarrow H_3^+ + D}$\\
D19 & $\mathrm{D + H_3^+ \rightarrow H_2D^+ + H}$ & D20 & $\mathrm{H_2D^+ + e^- \rightarrow H + H + D}$\\
D21 & $\mathrm{H_2D^+ + e^- \rightarrow H_2 + D}$ & D22 & $\mathrm{H_2D^+ + e^- \rightarrow HD + H}$\\
D23 & $\mathrm{H_2D^+ + H_2 \rightarrow H_3^+ + HD}$ & D24 & $\mathrm{H_2D^+ + H \rightarrow H_3^+ + D}$\\
\hline
\end{tabular}
\end{table}

\begin{table}[!h]
\caption{List of reactions for the helium chemistry. The rates are taken from \citet{GP98}.}
\label{tab:reache}
\centering
\scriptsize
\begin{tabular}{l l l l}
\hline
\hline
& Reaction & & Reaction\\
\hline
He1 & $\mathrm{He^{++} + e^- \rightarrow He^+ + \gamma}$ & He2 & $\mathrm{He^+ + \gamma \rightarrow He^{++} + e^-}$\\
He3 & $\mathrm{He^+ + e^- \rightarrow He + \gamma}$ & He4 & $\mathrm{He + \gamma \rightarrow He^+ + e^-}$\\
He5 & $\mathrm{He + H^+ \rightarrow He^+ + H}$ & He6 & $\mathrm{He^+ + H \rightarrow He + H^+}$\\
He7 & $\mathrm{He + H^+ \rightarrow HeH^+ + \gamma}$ & He8 & $\mathrm{He + H_2^+ \rightarrow HeH^+ + H}$\\
He9 & $\mathrm{He^+ + H \rightarrow HeH^+ + \gamma}$ & He10 & $\mathrm{HeH^+ + H \rightarrow He + H_2^+}$\\
He11 & $\mathrm{HeH^+ + e^- \rightarrow He + H}$ & He12 & $\mathrm{HeH^+ + H_2 \rightarrow H_3^+ + He}$\\
He13 & $\mathrm{HeH^+ + \gamma \rightarrow He + H^+}$ & He14 & $\mathrm{HeH^+ + \gamma \rightarrow He^+ + H}$\\
\hline
\end{tabular}
\end{table}

\begin{table}[!h]
\caption{List of reactions for the lithium chemistry. We consider the rates discussed in \citet{GP98}, except for the recombinations Li23 and Li25 \citep{ver96}, and for the photoionisations Li24 and Li26 \citep{ver96a}.}
\label{tab:reacli}
\centering
\scriptsize
\begin{tabular}{l l l l}
\hline
\hline
& Reaction & & Reaction\\
\hline
Li1 & $\mathrm{Li^+ + e^- \rightarrow Li + \gamma}$ & Li2 & $\mathrm{Li + \gamma \rightarrow Li^+ + e^-}$\\
Li3 & $\mathrm{Li^+ + H^- \rightarrow Li + H}$ & Li4 & $\mathrm{Li^- + H^+ \rightarrow Li + H}$\\
Li5 & $\mathrm{Li + e^- \rightarrow Li^- + \gamma}$ & Li6 & $\mathrm{Li^- + \gamma \rightarrow Li + e^-}$\\
Li7 & $\mathrm{Li + H^+ \rightarrow Li^+ + H}$ & Li8 & $\mathrm{Li + H^+ \rightarrow Li^+ + H + \gamma}$\\
Li9 & $\mathrm{Li + H^- \rightarrow LiH + e^-}$ & Li10 & $\mathrm{Li^- + H \rightarrow LiH + e^-}$\\
Li11 & $\mathrm{LiH^+ + H \rightarrow LiH + H^+}$ & Li12 & $\mathrm{LiH + H^+ \rightarrow LiH^+ + H}$\\
Li13 & $\mathrm{LiH + H \rightarrow Li + H_2}$ & Li14 & $\mathrm{Li^+ + H \rightarrow LiH^+ + \gamma}$\\
Li15 & $\mathrm{Li + H^+ \rightarrow LiH^+ + \gamma}$ & Li16 & $\mathrm{LiH + H^+ \rightarrow LiH^+ + H}$\\
Li17 & $\mathrm{LiH + H^+ \rightarrow Li^+ + H_2}$ & Li18 & $\mathrm{LiH^+ + e^- \rightarrow Li + H}$\\
Li19 & $\mathrm{LiH^+ + H \rightarrow Li + H_2^+}$ & Li20 & $\mathrm{LiH^+ + H \rightarrow Li^+ + H_2}$\\
Li21 & $\mathrm{LiH^+ + \gamma \rightarrow Li^+ + H}$ & Li22 & $\mathrm{LiH^+ + \gamma \rightarrow Li + H^+}$\\
Li23 & $\mathrm{Li^{2+} + e^- \rightarrow Li^+ + \gamma}$ & Li24 & $\mathrm{Li^+ + \gamma \rightarrow Li^{2+} + e^-}$\\
Li25 & $\mathrm{Li^{3+} + e^- \rightarrow Li^{2+} + \gamma}$ & Li26 & $\mathrm{Li^{2+} + \gamma \rightarrow Li^{3+} + e^-}$\\
\hline
\end{tabular}
\end{table}

\begin{table}[!h]
\caption{List of reactions for the carbon chemistry. The rates come from the \textsc{umist} database, except for five reactions: reaction C7 \citep{lip07}, the two charge transfers C8 and C9 between C, H and their respective ions \citep{sta98a}, the recombination C48 \citep{omu00} and the photoionisation C49, whose cross section is taken from \citet{ver96}.}
\label{tab:reacc}
\centering
\scriptsize
\begin{tabular}{l l l l}
\hline
\hline
& Reaction & & Reaction\\
\hline
C1 & C$^+$ + H$^-$ $\rightarrow$ H + C & C2 & $\mathrm{He^+ + C \rightarrow C^+ + He}$\\
C3 & $\mathrm{H^+ + C^- \rightarrow C + H}$ & C4 & $\mathrm{He^+ + C^- \rightarrow C + He}$\\
C5 & $\mathrm{C^+ + C^- \rightarrow C + C}$ & C6 & $\mathrm{C + e^- \rightarrow C^- + \gamma}$\\
C7 & $\mathrm{C^- + \gamma \rightarrow C + e^-}$ & C8 & $\mathrm{C^+ + H \rightarrow C + H^+}$\\
C9 & $\mathrm{C + H^+ \rightarrow C^+ + H}$ & C10 & $\mathrm{H + CH \rightarrow C + H_2}$\\
C11 & $\mathrm{H + CH \rightarrow C + H + H}$ & C12 & $\mathrm{C + H_2 \rightarrow CH + H}$\\
C13 & $\mathrm{H_2 + CH \rightarrow C + H + H_2}$ & C14 & $\mathrm{C^+ + H_2 \rightarrow CH^+ + H}$\\
C15 & $\mathrm{C + H \rightarrow CH + \gamma}$ & C16 & $\mathrm{H^+ + CH \rightarrow CH^+ + H}$\\
C17 & $\mathrm{H^- + C \rightarrow CH + e^-}$ & C18 & $\mathrm{H_2^+ + C \rightarrow CH^+ + H}$\\
C19 & $\mathrm{H_2^+ + CH \rightarrow CH^+ + H_2}$ & C20 & $\mathrm{H_3^+ + C \rightarrow CH^+ + H_2}$\\
C21 & $\mathrm{He^+ + CH \rightarrow C^+ + H + He}$ & C22 & $\mathrm{He^+ + CH \rightarrow CH^+ + He}$\\
C23 & $\mathrm{C^+ + H \rightarrow CH^+ + \gamma}$ & C24 & $\mathrm{C^+ + CH \rightarrow CH^+ + C}$\\
C25 & $\mathrm{CH^+ + H \rightarrow C^+ + H_2}$ & C26 & $\mathrm{CH^+ + e^- \rightarrow C + H}$\\
C27 & $\mathrm{H + C^- \rightarrow CH + e^-}$ & C28 & $\mathrm{H + CH_2 \rightarrow CH + H_2}$\\
C29 & H$_2$ + CH $\rightarrow$ CH$_2$ + H & C30 & $\mathrm{H^+ + CH_2 \rightarrow CH^+ + H_2}$\\
C31 & $\mathrm{H + CH_2^+ \rightarrow CH^+ + H_2}$ & C32 & $\mathrm{H_2^+ + CH \rightarrow CH_2^+ + H}$\\
C33 & $\mathrm{H_2 + CH^+ \rightarrow CH_2^+ + H}$ & C34 & $\mathrm{H_3^+ + CH \rightarrow CH_2^+ + H_2}$\\
C35 & $\mathrm{He^+ + CH_2 \rightarrow C^+ + He + H_2}$ & C36 & $\mathrm{He^+ + CH_2 \rightarrow CH^+ + He + H}$\\
C37 & $\mathrm{H^+ + CH_2 \rightarrow CH_2^+ + H}$ & C38 & $\mathrm{C + CH_2 \rightarrow CH + CH}$\\
C39 & $\mathrm{H_2^+ + CH_2 \rightarrow CH_2^+ + H_2}$ & C40 & $\mathrm{C^+ + CH_2 \rightarrow CH_2^+ + C}$\\
C41 & $\mathrm{CH_2^+ + e^- \rightarrow CH + H}$ & C42 & $\mathrm{CH_2^+ + e^- \rightarrow C + H + H}$\\
C43 & $\mathrm{CH_2^+ + e^- \rightarrow C + H_2}$ & C44 & $\mathrm{H^- + CH \rightarrow CH_2 + e^-}$\\
C45 & $\mathrm{H_2 + C^- \rightarrow CH_2 + e^-}$ & C46 & $\mathrm{H_2 + C \rightarrow CH_2 + \gamma}$\\
C47 & $\mathrm{H_2 + C^+ \rightarrow CH_2^+ + \gamma}$ & C48 & $\mathrm{C^+ + e^- \rightarrow C + \gamma}$\\
C49 & $\mathrm{C + \gamma \rightarrow C^+ + e^-}$ & &\\
\hline
\end{tabular}
\end{table}

\begin{table}[!h]
\caption{List of reactions for the nitrogen chemistry. We take the rates from the \textsc{umist} database, except for the photoionisation N41 \citep{ver96}.}
\label{tab:reacn}
\centering
\scriptsize
\begin{tabular}{l l l l}
\hline
\hline
& Reaction & & Reaction\\
\hline
N1 & H$^-$ + N$^+$ $\rightarrow$ N + H & N2 & C$^-$ + N$^+$ $\rightarrow$ N + C\\
N3 & CH + N$^+$ $\rightarrow$ N + CH$^+$ & N4 & H + NH $\rightarrow$ N + $\rm{H_2}$\\
N5 & $\rm{H_2}$ + N $\rightarrow$ NH + H & N6 & $\rm{H_{2}^{+}}$ + N $\rightarrow$ $\rm{NH^+}$ + H\\
N7 & $\rm{H_2}$ + N$^+$ $\rightarrow$ NH$^+$ + H & N8 & $\rm{H_2}$ + NH$^+$ $\rightarrow$ N + H$_{3}^{+}$\\
N9 & He$^+$ + NH $\rightarrow$ N$^+$ + He + H & N10 & H$^+$ + NH $\rightarrow$ NH$^+$ + H\\
N11 & H$_{2}^{+}$ + NH $\rightarrow$ NH$^+$ + H$_2$ & N12 & N$^+$ + NH $\rightarrow$ NH$^+$ + N\\
N13 & NH$^+$ + e$^-$ $\rightarrow$ N + H & N14 & H$^-$ + N $\rightarrow$ NH + e$^-$\\
N15 & CH + N $\rightarrow$ NH + C & N16 & N + $\rm{CH_2}$ $\rightarrow$ NH + CH\\
N17 & C + NH$^+$ $\rightarrow$ N + CH$^+$ & N18 & CH + NH$^+$ $\rightarrow$ CH$_{2}^{+}$ + N\\
N19 & C + NH $\rightarrow$ N + CH & N20 & H + NH$_2$ $\rightarrow$ NH + H$_2$\\
N21 & H$_2$ + NH $\rightarrow$ NH$_2$ + H & N22 & NH + NH $\rightarrow$ NH$_2$ + N\\
N23 & H$_{2}^{+}$ + NH $\rightarrow$ NH$_{2}^{+}$ + H & N24 & H$_2$ + NH$^+$ $\rightarrow$ NH$_{2}^{+}$ + H\\
N25 & H$_{3}^{+}$ + NH $\rightarrow$ NH$_{2}^{+}$ + H$_2$ & N26 & He$^+$ + NH$_2$ $\rightarrow$ N$^+$ + He + H$_2$\\
N27 & He$^+$ + NH$_2$ $\rightarrow$ NH$^+$ + He + H & N28 & NH$^+$ + NH $\rightarrow$ NH$_{2}^{+}$ + N\\
N29 & H$^+$ + NH$_2$ $\rightarrow$ NH$_{2}^{+}$ + H & N30 & H$_{2}^{+}$ + NH$_2$ $\rightarrow$ NH$_{2}^{+}$ + H$_2$\\
N31 & N$^+$ + NH$_2$ $\rightarrow$ NH$_{2}^{+}$ + N & N32 & NH$_{2}^{+}$ + e$^-$ $\rightarrow$ N + H + H\\
N33 & NH$_{2}^{+}$ + e$^-$ $\rightarrow$ NH + H & N34 & H$^-$ + NH $\rightarrow$ NH$_2$ + e$^-$\\
N35 & C + NH$_2$ $\rightarrow$ NH + CH & N36 & C + NH$_{2}^{+}$ $\rightarrow$ NH + CH$^+$\\
N37 & CH + NH$_{2}^{+}$ $\rightarrow$ NH + CH$_{2}^{+}$ & N38 & CH + NH$_{2}^{+}$ $\rightarrow$ NH$_2$ + CH$^+$\\
N39 & CH$_2$ + NH$_{2}^{+}$ $\rightarrow$ NH$_2$ + CH$_{2}^{+}$ & N40 & $\mathrm{N^+ + e^- \rightarrow N + \gamma}$\\
N41 & $\mathrm{N + \gamma \rightarrow N^+ + e^-}$ & &\\
\hline
\end{tabular}
\end{table}

\begin{table}[!h]
\caption{List of reactions for the oxygen chemistry. We consider the rates of the \textsc{umist} database, except for reactions O61 and O62 \citep{lebourlot1993}, for recombination O58 \citep{peq91} and for photoionisation O59, whose cross section is taken from \citep{ver96}.}
\label{tab:reaco}
\centering
\scriptsize
\begin{tabular}{l l l l}
\hline
\hline
& Reaction & & Reaction\\
\hline
O1 & O + e$^-$ $\rightarrow$ O$^-$ + $\gamma$ & O2 & H + O$^+$ $\rightarrow$ O + H$^+$\\
O3 & H$^+$ + O $\rightarrow$ O$^+$ + H & O4 & H$^-$ + O$^+$ $\rightarrow$ O + H\\
O5 & O$^-$ + $\gamma$ $\rightarrow$ O + e$^-$ & O6 & C$^-$ + O$^+$ $\rightarrow$ O + C\\
O7 & CH + O$^+$ $\rightarrow$ O + CH$^+$ & O8 & NH + O$^+$ $\rightarrow$ O + NH$^+$\\
O9 & H + O $\rightarrow$ OH + $\gamma$ & O10 & H + O$^-$ $\rightarrow$ OH + e$^-$\\
O11 & H + OH $\rightarrow$ O + H + H & O12 & H + OH $\rightarrow$ O + H$_2$\\
O13 & H$^-$ + O $\rightarrow$ OH + e$^-$ & O14 & H$^+$ + OH $\rightarrow$ OH$^+$ + H\\
O15 & H$_2$ + O $\rightarrow$ OH + H & O16 & H$_2$ + OH $\rightarrow$ O + H$_2$ + H\\
O17 & H$_2$ + O$^+$ $\rightarrow$ OH$^+$ + H & O18 & H$_{2}^{+}$ + O $\rightarrow$ OH$^+$ + H\\
O19 & H$_{2}^{+}$ + OH $\rightarrow$ OH$^+$ + H$_2$ & O20 & H$_{3}^{+}$ + O $\rightarrow$ OH$^+$ + H$_2$\\
O21 & O$^+$ + OH $\rightarrow$ OH$^+$ + O & O22 & OH$^+$ + e$^-$ $\rightarrow$ O + H\\
O23 & He$^+$ + OH $\rightarrow$ O$^+$ + He + H & O24 & C + OH $\rightarrow$ O + CH\\
O25 & C + OH$^+$ $\rightarrow$ O + CH$^+$ & O26 & CH + O $\rightarrow$ OH + C\\
O27 & CH + OH$^+$ $\rightarrow$ OH + CH$^+$ & O28 & OH + F $\rightarrow$ HF + O\\
O29 & N + OH $\rightarrow$ O + NH & O30 & N$^+$ + OH $\rightarrow$ OH$^+$ + N\\
O31 & NH + O $\rightarrow$ OH + N & O32 & NH$^+$ + O $\rightarrow$ OH$^+$ + N\\
O33 & H$_2$O$^+$ + e$^-$ $\rightarrow$ O + H + H & O34 & H$_2$O$^+$ + e$^-$ $\rightarrow$ O + H$_2$\\
O35 & H$_2$O$^+$ + e$^-$ $\rightarrow$ OH + H & O36 & O + H$_2$O $\rightarrow$ OH + OH\\
O37 & O$^+$ + H$_2$O $\rightarrow$ H$_2$O$^+$ + O & O38 & H$_2$ + O$^-$ $\rightarrow$ H$_2$O + e$^-$\\
O39 & H$_2$ + H$_2$O $\rightarrow$ OH + H$_2$ + H & O40 & H$_2$ + OH $\rightarrow$ H$_2$O + H\\
O41 & H$_2$ + OH$^+$ $\rightarrow$ H$_2$O$^+$ + H & O42 & H$_{2}^{+}$ + OH $\rightarrow$ H$_2$O$^+$ + H\\
O43 & H$_{3}^{+}$ + O $\rightarrow$ H$_2$O$^+$ + H & O44 & H$_{3}^{+}$ + OH $\rightarrow$ H$_2$O$^+$ + H$_2$\\
O45 & H + H$_2$O $\rightarrow$ OH + H$_2$ & O46 & H + H$_2$O $\rightarrow$ OH + H + H\\
O47 & H + OH $\rightarrow$ H$_2$O + $\gamma$ & O48 & H$^-$ + OH $\rightarrow$ H$_2$O + e$^-$\\
O49 & He$^+$ + H$_2$O $\rightarrow$ OH + He + H$^+$ & O50 & He$^+$ + H$_2$O $\rightarrow$ OH$^+$ + He + H\\
O51 & OH + OH $\rightarrow$ H$_2$O + O & O52 & OH$^+$ + OH $\rightarrow$ H$_2$O$^+$ + O\\
O53 & OH$^+$ + H$_2$O $\rightarrow$ H$_2$O$^+$ + OH & O54 & H$_2$O + F $\rightarrow$ HF + OH\\
O55 & C + H$_2$O$^+$ $\rightarrow$ OH + CH$^+$ & O56 & NH + OH $\rightarrow$ H$_2$O + N\\
O57 & NH$^+$ + OH $\rightarrow$ H$_2$O$^+$ + N & O58 & $\mathrm{O^+ + e^- \rightarrow O + \gamma}$\\
O59 & $\mathrm{O + \gamma \rightarrow O^+ + e^-}$ & O60 & $\mathrm{H^+ + H_2O \rightarrow H_2O^+ + H}$\\
O61 & $\mathrm{OH^+ + H \rightarrow O^+ + H_2}$ & O62 & $\mathrm{H_2O^+ + H \rightarrow H^+ + H_2O}$\\
O63 & $\mathrm{H_2^+ + H_2O \rightarrow H_2O^+ + H_2}$ & &\\
\hline
\end{tabular}
\end{table}

\begin{table}[!h]
\caption{List of reactions for the fluorine chemistry. We take the rates discussed in \citet{puy07}.}
\label{tab:reacf}
\centering
\scriptsize
\begin{tabular}{l l l l}
\hline
\hline
& Reaction & & Reaction\\
\hline
F1 & F + H $\rightarrow$ HF + $\gamma$ & F2 & FH + $\gamma$ $\rightarrow$ F + H\\
F3 & F + e$^-$ $\rightarrow$ F$^-$ + $\gamma$ & F4 & F$^-$ + $\gamma$ $\rightarrow$ F + e$^-$\\
F5 & F$^-$ + H $\rightarrow$ HF + e$^-$ & F6 & HF + e$^-$ $\rightarrow$ F$^-$ + H\\
F7 & F + H$_2$ $\rightarrow$ HF + H & &\\
\hline
\end{tabular}
\end{table}

\end{appendix}

\end{document}